\documentclass[twocolumn]{aastex62}     

\usepackage{amsmath,amsfonts,amssymb}
\usepackage{graphicx}
\usepackage{multirow}
\usepackage{ulem}
\usepackage{tablefootnote}
\usepackage{threeparttable}
\usepackage{natbib}
\submitjournal{ApJS}   

\shorttitle{A comparison of calibration techniques}  
\shortauthors{Maud et al.}

\begin{document}

\title{
    ALMA High-frequency Long-baseline Campaign in 2017:  \\
    A Comparison of the Band-to-band and In-band Phase Calibration Techniques and\\
    Phase-calibrator Separation Angles}

\correspondingauthor{Luke T. Maud}
\email{lmaud@eso.org}

\author[0000-0002-7675-3565]{Luke T. Maud}
\affil{ESO Headquarters, 
        Karl-Schwarzchild-Str 2 85748 Garching, Germany}
\affil{Allegro, Leiden Observatory, Leiden University, 
        PO Box 9513, 2300 RA Leiden, The Netherlands} 
\nocollaboration{}

\author[0000-0002-0976-4010]{Yoshiharu Asaki}
\affil{Joint ALMA Observatory, 
        Alonso de C\'{o}rdova 3107, Vitacura, Santiago, 763 0355, Chile}
\affil{National Astronomical Observatory of Japan, \\
        Alonso de C\'{o}rdova 3788, Office 61B, Vitacura, Santiago, Chile}
\affil{Department of Astronomical Science, School of Physical Sciences, \\
        The Graduate University for Advanced Studies (SOKENDAI), 
        2-21-1 Osawa, Mitaka, Tokyo 181-8588, Japan}
\nocollaboration{}        

\author{Edward B. Fomalont}
\affil{Joint ALMA Observatory, 
        Alonso de C\'{o}rdova 3107, Vitacura, Santiago, 763 0355, Chile}
\affil{National Radio Astronomy Observatory, 
        Edgemont Rd. Charlottesville, VA 22903, USA}
\nocollaboration{}

\author{William R. F. Dent}
\affil{Joint ALMA Observatory, 
        Alonso de C\'{o}rdova 3107, Vitacura, Santiago, 763 0355, Chile}
\nocollaboration{}

\author{Akihiko Hirota}
\affil{Joint ALMA Observatory, 
        Alonso de C\'{o}rdova 3107, Vitacura, Santiago, 763 0355, Chile}
\affil{National Astronomical Observatory of Japan, \\
        Alonso de C\'{o}rdova 3788, Office 61B, Vitacura, Santiago, Chile}
\nocollaboration{}

\author{Satoki Matsushita}
\affil{Institute of Astronomy and Astrophysics, Academia Sinica,\\
   11F of Astronomy-Mathematics Building, AS/NTU,\\
   No.1, Sec. 4, Roosevelt Rd, Taipei 10617, Taiwan, R.O.C.}
   \nocollaboration{}

\author{Neil M. Phillips}
\affil{ESO Headquarters, 
        Karl-Schwarzchild-Str 2 85748 Garching, Germany}
\nocollaboration{}

\author{John M. Carpenter}
\affil{Joint ALMA Observatory, 
        Alonso de C\'{o}rdova 3107, Vitacura, Santiago, 763 0355, Chile}
\nocollaboration{}

\author{Satoko Takahashi}
\affil{Joint ALMA Observatory, 
        Alonso de C\'{o}rdova 3107, Vitacura, Santiago, 763 0355, Chile}
\affil{National Astronomical Observatory of Japan, \\
        Alonso de C\'{o}rdova 3788, Office 61B, Vitacura, Santiago, Chile}
\affil{Department of Astronomical Science, School of Physical Sciences, \\
        The Graduate University for Advanced Studies (SOKENDAI), 
        2-21-1 Osawa, Mitaka, Tokyo 181-8588, Japan}
\nocollaboration{}

\author{Eric Villard}
\affil{Joint ALMA Observatory, 
        Alonso de C\'{o}rdova 3107, Vitacura, Santiago, 763 0355, Chile}
\nocollaboration{}

\author{Tsuyoshi Sawada}
\affil{Joint ALMA Observatory, 
        Alonso de C\'{o}rdova 3107, Vitacura, Santiago, 763 0355, Chile}
\affil{National Astronomical Observatory of Japan, \\
        Alonso de C\'{o}rdova 3788, Office 61B, Vitacura, Santiago, Chile}
\nocollaboration{}

\author{Stuartt Corder}
\affil{Joint ALMA Observatory, 
        Alonso de C\'{o}rdova 3107, Vitacura, Santiago, 763 0355, Chile}
\nocollaboration{}


\begin{abstract}
The Atacama Large millimeter/submillimeter Array (ALMA) obtains spatial resolutions of 15 to 5 milli-arcsecond (mas) at 275-950\,GHz (0.87-0.32\,mm) with 16\,km baselines. Calibration at higher-frequencies is challenging as ALMA sensitivity and quasar density decrease. The Band-to-Band (B2B) technique observes a detectable quasar at lower frequency that is closer to the target, compared to one at the target high-frequency. Calibration involves a nearly constant instrumental phase offset between the frequencies and the conversion of the temporal phases to the target frequency. The instrumental offsets are solved with a differential-gain-calibration (DGC) sequence, consisting of alternating low and high frequency scans of strong quasar. Here we compare B2B and in-band phase referencing for high-frequencies ($>$289\,GHz) using 2-15\,km baselines and calibrator separation angles between $\sim$0.68 and $\sim$11.65$^{\circ}$. The analysis shows that: (1) DGC for B2B produces a coherence loss $<$7\,\% for DGC phase RMS residuals $<$30$^{\circ}$. (2) B2B images using close calibrators ( $<$1.67$^{\circ}$ ) are superior to in-band images using distant ones ( $>$2.42$^{\circ}$ ). (3) For more distant calibrators, B2B is preferred if it provides a calibrator $\sim$2$^{\circ}$ closer than the best in-band calibrator. (4) Decreasing image coherence and poorer image quality occur with increasing phase calibrator separation angle because of uncertainties in the antenna positions and sub-optimal phase referencing. (5) To achieve $>$70\,\% coherence for long-baseline (16 km) band 7 (289GHz) observations, calibrators should be within $\sim$4$^{\circ}$ of the target.
\end{abstract}

\keywords{Long baseline interferometry (932), Submillimeter astronomy (1647), Phase error (1220)}

\section{Introduction}
\label{intro}

ALMA is currently the only submillimeter interferometer that provides access to frequencies between 420$-$950\,GHz, wavelengths 0.3$-$0.7\,mm \citep{Baryshev2015,Gonzalez2014}. Theoretically, ALMA can achieve a resolution as fine as $\sim$5\,mas if observing at 950\,GHz (i.e. ALMA band 10) using the maximal baselines of 16\,km. Importantly, this resolution translates into sub-au scales for sources located $<$200\,pc away, such as protoplanetary discs, or sub-pc scales for extra-galactic sources within 40\,Mpc. The initial ALMA long-baseline observations were originally showcased in \citep{ALMA2015a,ALMA2015b,ALMA2015c} and have been offered as an observing mode for band 3, 4 and 6 since 2015, and in band 5 shortly thereafter. Starting in ALMA Cycle 7, band 7 long-baselines have been offered for use, where resolutions can reach $\sim$15\,mas. However, observations at frequencies higher than 450\,GHz have not been offered yet on baselines longer than 5\,km. These observations pose a significant challenge because standard calibration techniques are more difficult to employ.

Observations in the sub-mm regime suffer from absorption, in that a signal from an astronomical source is attenuated, mostly by water vapor, in the Earth's atmosphere. These signals cannot be recovered and observations must be limited to conditions with low precipitable water vapor (PWV) content to maximize transmission. ALMA typically limits band 9 and band 10 observations to when the precipitable water vapor is less than $\sim$0.66\,mm and $\sim$0.47\,mm, respectively. These conditions occur only $\sim$20-25\,\% of the time on the Chajnantor plateau where ALMA is located. Moreover, interferometric observations rely on the coherence of signals for all baseline in order to successfully image a scientific target. Fluctuations in the troposphere introduce spatial and temporal variable delays in the path length of these signals. These path length variations ($\ell$, in $\mu$m) directly relate to root-mean-squared (RMS) of the phase fluctuations ($\sigma_{\phi}$, in radians) for a given observing frequency ($\nu_{obs}$ in Hz) by:

\begin{equation}
  \label{eqn0}
\sigma_{\phi} = 2\pi\ell \, \frac{\nu_{obs}}{c} \quad(radians),
\end{equation}

\noindent where $c$ is the speed of light (in $\mu$m\,s$^{-1}$). Thus for particular atmospheric conditions, causing path length variations, the phase fluctuations will increase with increasing observing frequency\footnote{This is generally true although dispersion can occur near atmospheric lines.}. The estimated coherence\footnote{We use the term `estimated coherence' throughout this paper to denote coherence values estimated from Equation\,\ref{eqn1} after inputting a measured phase RMS fluctuation and ${\bf V}_0$ = 1.} for the visibilities (${\bf V} = {\bf V}_0e^{i\phi}$, where ${\bf V}_0$ are the true visibilities and $\phi$ describes the phase fluctuations caused by the atmosphere, \citealt{Thompson2017}) can be calculated using:

\begin{equation}
  \label{eqn1}
\langle {\bf V} \rangle = \langle {\bf V}_0 \rangle \, \langle e^{i\phi} \rangle = \langle {\bf V}_0 \rangle \, e^{-\sigma^2_{\phi}/2},
\end{equation}

\noindent assuming Gaussian random phase fluctuations, $\phi$, with an RMS of $\sigma_{\phi}$ (in radians) about a mean phase of zero \citep{Thompson2017}. Anomalous path length changes on many baselines can cause a significant loss in the signal due to decoherence and also cause blurring and smearing in target image \citep[e.g.][]{Carilli1999}.

Atmospheric fluctuations are thought to be described by Kolmogorov turbulence theory \citep{Coulman1990} where path length variations are a function of baseline length. During previous ALMA long-baseline campaigns, \citet{Matsushita2017} indicated that the phase RMS $\sigma_{\phi}$ increases as $b^{0.6}$ for baselines, $b$, $<$1\,km and as $b^{0.2}$ for $b >$1\,km. Thus, long-baseline observations are the most susceptible to large phase fluctuations which much be corrected in order to achieve high coherence imaging.

ALMA employs a system of water-vapor-radiometers (WVRs) that measure variations in the 183\,GHz water line profile directly in the line-of-sight of each 12\,m antenna. These variations can be translated into a path length and therefore used to correct for tropospheric phase variations that are predominantly caused by water vapor \citep[e.g.][]{Lay1997,Delgado2000,Nikolic2007,Nikolic2012, Nikolic2013,Stirling2005, Maud2017}. Typically the effective path length variations are reduced by a factor of 2 when the PWV is $>$1\,mm, but the effect is reduced in dry conditions where the PWV is $<$1\,mm \citep{ALMA2015,Matsushita2017,Maud2017}. When the PWV is $<$0.7-0.8\,mm, as required for the higher frequency observations ($>$420\,GHz), the WVRs will not provide a significant improvement of the phases as the water vapor fluctuations are no longer dominant. 

Phase referenced observations are the standard for most interferometric observations. A point source calibrator, typically a quasar, close-by on the sky is observed interspersed with the observations of the astronomical source to form a phase referencing cycle. The phase referencing cycle, from calibrator scan to target scan and back to the calibrator scan, is repeated with a finite cycle time. During data calibration the phase solutions are interpolated and transferred from the calibrator scans to the source scans, correcting any phase variability occurring on timescales longer than the referencing cycle time. 
In order to counteract rapid atmospheric variations fast-switching, with notably reduced cycle times, can be used. Pre-ALMA-era investigations found that cycle times as low as 80\,s lead to better calibration and thus better imaging \citep [e.g.][]{Holdaway1995,Carilli1997,Morita2000,Lal2007}. These authors also noted that even faster cycling might be required for high frequency observations. For ALMA, the antennas can change pointing by a few degrees in only 2-3\,s such that fast-switching phase referencing cycle times can be as short as $\sim$20\,s without significant overheads \citep{Asaki2014,Asaki2016}. 

For completeness, we note there are special cases when self-calibration can be undertaken. If the science target is a sufficiently strong source and has a high surface brightness it can itself be used to calibrate the phase delays on the timescales of those variations (for more details see \citealt{Pearson1984}, \citealt{Cornwell1999}, and \citealt{Brogan2018}). Self-calibration requires an initial model of the source brightness distribution which usually remains unknown prior to the observations. Unless a point-like initial starting model can be used, phase referencing needs to be good enough to provide the starting image for self-calibration.

For most high-frequency long-baseline observations, we need relatively short cycle times and a close calibrator to track the phase variations. The requirement of a close calibrator is compounded by path length errors caused by antenna position uncertainties. \citet{Hunter2016} present a fit to ALMA long-baseline antenna position measurements indicating uncertainties of up to $\sim$0.2\,mm/km, part of which is caused by unknown tropospheric delays over the array. For ALMA long-baselines, the inner product of the baseline uncertainty vector and the separation angle vector between a science target and phase calibrator \citep{Asaki2020} can lead to a non-negligible path length error dependent on both baseline length and frequency. Thus, a distant phase calibrator would impart an incorrect phase solution to the target source. 

A major problem is that quasars used as phase calibrators have a negative spectral index ($\alpha \sim$ -0.7) and are notably weaker above 420\,GHz, while atmospheric transmission decreases and reduces observing sensitivity. Weaker calibrators require more integration time to achieve enough signal-to-noise to provide a single calibration solution. This entirely conflicts with the necessity of short cycle times to track the variable atmosphere as there would be little time left to observe a target source. A more subtle point is that long on-calibrator times should be avoided and cannot be averaged as the signal itself would suffer from decorrelation due to the fluctuations that we are trying to correct. Alternatively, one could search for bright enough calibrators with increasing distance from a science target. Studies by \citet{Asaki1998,Asaki2016} however, indicated that using distant calibrators does not always provide the expected levels of phase correction (see also Section \ref{hardlim}).

One foreseeable way to calibrate high-frequency long-baseline observations is to observe the calibrators at a lower frequency where they are naturally much stronger. Subsequently, the likelihood of finding a calibrator close to a given scientific target source improves \citep{Asaki2020} and hence would minimize any phase errors caused due to large calibrator separation angles. Because phase variations scale linearly with frequency (Equation \ref{eqn0}), phase solutions can be scaled by a multiplicative factor from a calibrator observed at a low frequency to a scientific target observed at a higher frequency. Transferring phase solutions from lower to higher frequencies has being tried, somewhat successfully, by other arrays: the Nobeyama millimeter array (NMA) used paired antennas, one set observing a satellite at 19.45\,GHz, while the others observed a celestial target at 146.81\,GHz \citep{Asaki1998}; The Combined Array for Research in Millimeter-wave (CARMA) also used a paired-antenna-combination-system (C-PACS) in which their long-baseline (2\,km) antennas were paired with adjacent antennas that observed at a wavelength of 1\,cm \citep{Perez2010a,Perez2010b,Zauderer2016}; the Sub-Millimeter-Array (SMA) tested a dual-frequency mode in which two frequencies were observed simultaneously, and the phase solution derived from the low-frequency observations were applied to the high-frequency data \citep{Hunter2005}. The SMA, before ALMA, was the only interferometric instrument to attempt frequencies as high 650\,GHz, as ALMA can now observe. The SMA observations were very difficult even with the dual frequency capability as the culmination of atmospheric conditions, lower sensitivities (a smaller array with smaller antennas when compared with ALMA) and instabilities meant that very few successful studies were undertaken. In the dual-frequency observations by \citet{Chen2007} low- to high-frequency phase transfer was not used, rather calibration relied on solutions transferred from a high-frequency maser source $\sim$5$^{\circ}$ away.

Currently, the KVN (Korean-Very-Long-Baseline-Interferometry-Network) can simultaneously observe at 22, 43, 86 and 129\,GHz and is the only instrument to regularly employ a technique termed frequency-phase-transfer (FPT - \citealt{Rioja2015}), which is the process of calibrating all sources at a given high-frequency with the lower-frequency solutions. Note, these are VLBI observations of point-like sources at cm-wavelengths such that the FPT is valid. An additional technique, source-frequency-phase-referencing (SFPR - \citealt{Dodson2009, Rioja2011}) can also be employed. This uses a combination of FPT with simultaneous multi-band phase referencing allowing a scaling of lower- to higher-frequency phase solutions to correct tropospheric phase variations while the phase referencing corrects slower ionospheric errors.

For ALMA the technique analogous to FPT is called band-to-band (B2B)\footnote{Different ALMA frequency receiver ranges are divided into bands.} phase referencing and was already envisaged as a calibration method for ALMA combined with fast switching \citep[e.g.][]{Holdaway2004}. The use case of B2B would be that a phase calibrator at a lower frequency would generally be found closer to a selected science target compared to using standard in-band phase referencing. We investigate this scenario in detail in this paper. \\

In this paper, as part of the series on the HF-LBC-2017 \citep[see also][]{Asaki2020,Asaki2020b,Maud2020}, we make comparisons of the B2B and the standard in-band phase referencing techniques. These comparisons are part of the stage 3 tests enumerated in \citet{Asaki2020}. For stage 3 there are three main goals that can be investigated with our observations, of which the first two are detailed in this work whereas goal (3) is detailed in \citet{Maud2020}:
\begin{enumerate}
\item  To make a comparison of the image quality obtained from B2B phase referencing with that of in-band calibration using the same phase calibrator and to categorise any detrimental effects due to differential-gain-calibration (DGC).
\item To determine the deterioration of the standard in-band phase referencing with increasing calibrator distance from the target, and to contrast with B2B observations using closer calibrators.
\item To determine the image improvement as a function of phase referencing cycle time from $\sim$120 to 24\,s. 
\end{enumerate}

In Section \ref{obs} we describe in detail the observational tests, the data reduction and the methodology used for the analysis. Section \ref{res} details the results of the first two main goals listed above and the comparisons between in-band and B2B techniques. In Section \ref{disc} we specifically discuss the effects of calibrator separation angle and the choice of observing technique, B2B or in-band. Finally, we summarize our findings in Section \ref{sum}. 

\section{Observations, reduction and methodology}
\label{obs}
We undertook our experiments in the latter half of Cycle 4 and the start of Cycle 5 (2017 June to October) as part of the 2017 high-frequency and long-baseline campaign HF-LBC-2017 \citep{Asaki2020}. During this period of time we conducted 50 full length observations, of which 44 are useable for analysis\footnote{One band 10 test and four band 9 tests failed due to non-detections as a result of high image noise; while one band 7 could not be calibrated due to missing calibrator scans.}. Spreading the tests over a period of a few months allowed a reasonable coverage of maximal baseline lengths ranging from $\sim$2\,km out to the longest baselines in September and October of $\sim$15\,km. Observations were taken using a range of band-to-band pairs (i.e. frequency pairs) B7-3, B8-4, B9-4, B9-6 and even one B10-7 observation (where the first number denotes the target frequency band and the latter the calibrator frequency band). The frequency pairs are constrained by the ALMA hardware \citep[see][]{Asaki2020}. The higher band of the pairings were used alone for in-band observations, B7, B8, B9 and B10. 

Suitable quasars were selected as targets and calibrators to cover a variety of local sidereal time (LST) ranges to provide some coverage of different stability conditions (from early evening, through the night and into the early morning). We also minimized the impact on ALMA science observations by conducting tests in undersubscribed regions of the schedule or when too few antennas were available to meet the operations criteria. The number of antennas available for our tests ranged from 13 to 48, although not all observations with low antenna numbers were usable. We did not impose strict phase RMS stability constraints before triggering the tests, so that a wide range of atmospheric conditions could be explored.

\begin{figure*}  
\begin{center}
  \includegraphics[width=15.5cm]{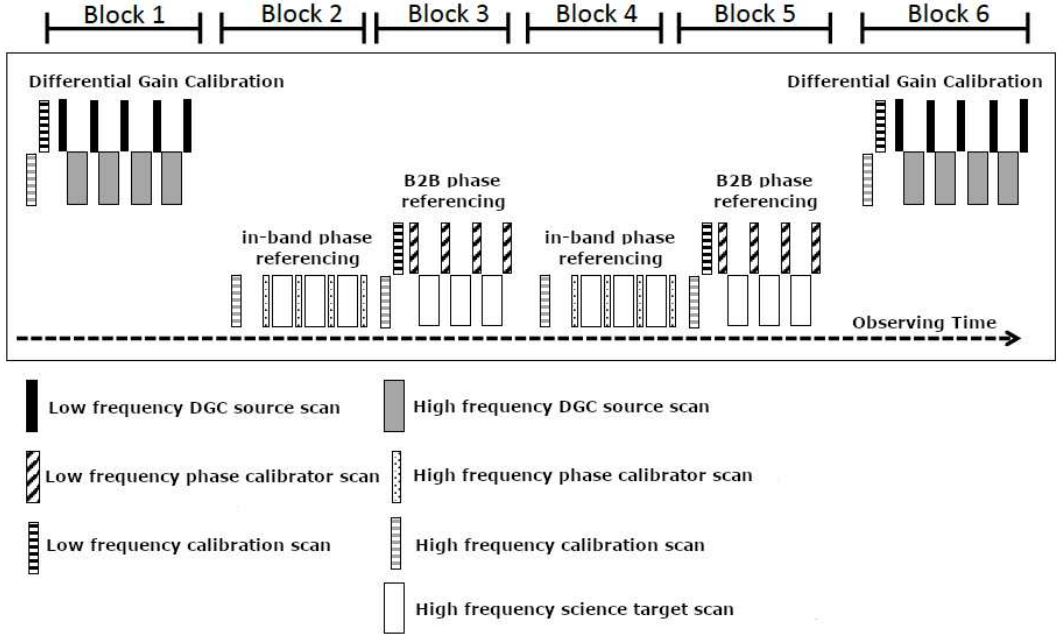}  
\caption{Illustrative schematic of the full observing sequence. The six blocks making up one full observation can be understood by separating the first DGC observation, the first in-band referencing, the first B2B referencing, the second in-band block, the second B2B block, and the end DGC block. The figure is illustrative as it is not accurate in terms of observing time or number of scans. In reality all scan lengths were fixed at 9\,s for all targets regardless of observing at the high or low frequency, and there were more than 10 scans per block. The individual scans are identified in the legend at the bottom. Calibration scans refer to system-noise temperature (Tsys) and pointing scans.}
\label{fig1} 
\end{center}
\end{figure*}

Each observation consisted of six blocks, as shown in Figure \ref{fig1}. In the first block the DGC source (see Section \ref{dgc_sec} and \citealt{Asaki2020,Asaki2020b}) is used to measure the instrumental phase difference between the high- and low-frequencies. Blocks 2,3,4,5 alternate between normal in-band phase referencing and B2B. Block 6 again targets the DGC source. Due to the available telescope software in 2017 the six blocks produced six individual datasets that were later combined. In general a full sequence ran for approximately 50\,minutes including all required system temperature and pointing calibrations. Typically $\sim$4\,minutes were spent on the target source for each of the in-band and B2B observations respectively. Table \ref{tab0} gives an overview of the 44 analyzed experiments, including the maximal baseline lengths and time of day.

To address the first goal, some observations use the same calibrator for B2B and for in-band ($<$2$^{\circ}$), such that we can investigate any loss of image quality since the DGC potentially adds extra phase errors for B2B. For the second goal, to illustrate the real benefit of B2B calibration when the best in-band calibration is much farther away, we use close calibrators for B2B ($<$2$^{\circ}$) and more distant ones for in-band (2-11$^{\circ}$). The phase referencing cycle time (i.e. measured from the mid-point of the phase calibration scan to the mid-point of the subsequent phase calibrator scan) was of the order 24\,s both for the in-band and B2B phase referencing blocks. We use fast-switching when compared to the standard ALMA Cycle 7 long-baseline cycle-time of 72\,s. The integration time used was 1.15\,s, while each on-source scan comprised eight integrations and thus was 9.2\,s in length (i.e. from the calibrator mid-point, 4.6\,s calibrator $+$ 2-3\,s slew/freq.\,change $+$ 9.2\,s target $+$ 2-3\,s slew/freq.\,change $+$ 4.6\,s calibrator $\sim$ 24\,s cycle time). As noted, the antennas can slew between sources and stabilize in 2-3 seconds. Specifically for the B2B blocks, short cycle times are possible only by changing frequency using the harmonic frequency switching mode \citep[see also][]{Asaki2020}. Briefly, we use a fixed frequency for the photonic Local Oscillator (LO) that is tuned once at the start of the observations. Each receiver multiplies this LO value using an auxiliary oscillator in each antenna to achieve the first LO frequency (LO1) and set the frequency ranges. The harmonic frequency switching mode permits a change in frequency in 2-3 seconds. In comparison any normal frequency changes at ALMA where both the LO and LO1 are reset requires a 20\,s overhead to re-lock the frequencies and thus would not have allowed the fast switching B2B operation. All observations were performed in the continuum TDM (Time Domain Mode) for the correlator setup. In this mode there are four spectral windows (SPWs) per frequency, configured with 64 channels. The bandwidth of each SPW is 2\,GHz, with a usable width of 1.875\,GHz, providing an aggregate effective bandwidth of 7.5\,GHz. In the case of the band 9 observations, due to the specific testing setup, only two of the four SPWs are independent providing an effective bandwidth of 3.75\,GHz. The central frequencies for the observations are listed in Table \ref{tab1}. The high-frequencies are those of the target and calibrator for in-band, while the low frequencies are those for the calibrator only when using B2B mode.

\begin{table*}[h!]
\begin{center}
  \caption{Overview and parameters of the 44 analyzed in-band - B2B observations conducted as part of the HF-LBC-2017.} 
  {\scriptsize  
\begin{tabular}{@{}lllrrrrrrrr@{}}
\hline
Name & Date & Time  & \multicolumn{3}{c}{DGC source}    & No.   &  PWV  & \multicolumn{2}{c}{Expected phase RMS} & Maximum \\ 
 &  & (UTC)  &   Name  &  flux (Jy) & $\alpha$  & ants  &  (mm)  & ($\mu$m) & (deg) & Baseline (m) \\
\hline
\hline
\multicolumn{11}{c}{\bf{Band 7 - 3}} \\   
\hline
\hline
J2228-170827-B73-1deg&2017-08-27&04:48:24&J2253+1608&5.89&-0.705&48&0.62&35.4&10.9&4540\\
J2228-170829-B73-1deg&2017-08-29&07:33:25&J2253+1608&5.89&-0.705&48&1.45&42.0&14.1&4734\\
J2228-170829-B73-3deg&2017-08-29&02:07:47&J2253+1608&5.89&-0.705&47&1.33&55.5&18.6&5049\\
J2228-170829-B73-6deg&2017-08-29&02:57:20&J2253+1608&5.89&-0.705&47&1.28&41.6&13.9&4874\\
J0449-170829-B73-2deg&2017-08-29&08:24:35&J0522$-$3627&4.44&-0.311&48&1.57&38.8&13.0&4798\\
J0449-170829-B73-3deg&2017-08-29&09:21:55&J0522$-$3627&4.44&-0.311&48&1.70&42.2&14.1&4807\\
J2228-170830-B73-1deg&2017-08-30&03:12:06&J2253+1608&5.98&-0.690&45&2.90&50.0&16.7&4790\\
J2228-170830-B73-3deg&2017-08-30&04:00:44&J2253+1608&5.98&-0.690&45&2.51&45.8&15.3&4531\\
J0449-170830-B73-5deg$^a$&2017-08-30&08:03:25&J0522$-$3627&4.44&-0.311&47&2.43&45.2&15.2&4783\\ 
J0449-170830-B73-7deg&2017-08-30&08:57:09&J0522$-$3627&4.44&-0.311&47&2.40&33.6&11.2&4799\\
J2228-170917-B73-1deg&2017-09-17&01:45:48&J2253+1608&5.41&-0.669&41&1.84&59.7&20.0&12115\\
J0633-170917-B73-1deg&2017-09-17&14:05:08&J0522$-$3627&4.04&-0.252&39&1.61&153.8&51.5&11373\\
J0633-170917-B73-4deg&2017-09-17&14:52:56&J0522$-$3627&4.04&-0.252&39&1.57&270.0&90.4&11366\\
J2228-170926-B73-1deg&2017-09-26&03:09:05&J2253+1608&4.94&-0.688&42&0.97&30.2&10.1&13916\\
J2228-170926-B73-3deg$^b$&2017-09-26&03:47:41&J2253+1608&4.94&-0.688&42&1.05&36.5&12.2&13015\\
J2228-170926-B73-6deg&2017-09-26&04:38:23&J2253+1608&4.94&-0.688&42&1.04&35.2&11.8&11457\\
J0449-170928-B73-2deg&2017-09-28&09:02:33&J0522$-$3627&4.19&-0.179&43&0.52&51.0&17.1&14818\\
J0449-170929-B73-5deg&2017-09-29&07:05:54&J0522$-$3627&3.95&-0.179&42&1.27&88.8&29.7&14694\\
J0633-170930-B73-1deg$^{c,d,e}$&2017-09-30&09:41:19&J0522$-$3627&3.95&-0.179&48&0.83&22.2&7.4&14951\\
J0633-170930-B73-4deg$^{e,f}$&2017-09-30&08:51:18&J0522$-$3627&3.95&-0.179&48&0.83&22.2&7.4&14961\\ 
J0633-171001-B73-9deg$^{c}$&2017-10-01&09:25:16&J0522$-$3627&3.95&-0.179&40&1.61&239.8&80.3&14470\\
\hline
\hline
\multicolumn{11}{c}{\bf{Band 8 - 4}} \\ 
\hline
\hline
J1709-170717-B84-2deg&2017-07-17&00:49:17&J1924$-$2914&2.68&-0.583&26&0.65&72.1&34.1&2058\\
J1709-170717-B84-1deg&2017-07-17&01:45:23&J1924$-$2914&2.68&-0.583&26&0.66&57.2&27.0&2058\\
J1709-170717-B84-11deg&2017-07-17&02:32:47&J1924$-$2914&2.68&-0.583&26&0.64&69.0&32.6&2055\\
J2228-170717-B84-7deg$^{c,g}$&2017-07-17&09:58:57&J2253+1608&4.98&-0.716&28&0.26&105.8&50.0&2249\\  
J1259-170717-B84-1deg&2017-07-17&22:40:09&J1256$-$0547&6.72&-0.495&35&0.81&52.5&24.8&3447\\
J1259-170717-B84-8deg&2017-07-17&23:30:44&J1256$-$0547&6.72&-0.495&35&0.76&48.3&22.8&3327\\
J1259-170718-B84-11deg&2017-07-18&00:22:30&J1256$-$0547&6.72&-0.495&35&0.74&61.6&29.2&3229\\
J0633-170718-B84-1deg&2017-07-18&13:52:33&J0522$-$3627&5.11&-0.148&34&0.34&35.8&16.9&3649\\
J0633-170718-B84-4deg$^h$&2017-07-18&14:41:55&J0522$-$3627&5.11&-0.148&34&0.41&30.7&14.5&3688\\
J0633-170718-B84-9deg&2017-07-18&16:48:00&J0522$-$3627&5.11&-0.148&34&0.59&46.6&22.1&3696\\
J0633-170718-B84-6deg&2017-07-18&15:55:11&J0522$-$3627&5.11&-0.148&34&0.52&51.7&24.5&3691\\
J2228-170819-B84-1deg$^i$&2017-08-19&08:01:17&J2253+1608&5.34&-0.665&43&0.66&39.8&18.8&3290\\
J2228-170820-B84-3deg&2017-08-20&04:26:10&J2253+1608&4.62&-0.705&24&0.72&107.8&51.0&5016\\
J2228-170820-B84-10deg&2017-08-20&05:12:57&J2253+1608&4.62&-0.705&24&0.67&90.5&42.8&5322\\
J2228-170820-B84-1deg&2017-08-20&06:08:47&J2253+1608&4.62&-0.705&24&0.59&88.8&42.0&5357\\
\hline
\hline
\multicolumn{11}{c}{\bf{Band 9 - 4}} \\
\hline
\hline
J2228-170717-B94-1deg&2017-07-17&06:58:47&J2253+1608&3.37&-0.716&29&0.35&40.56&33.2&2343\\ 
\hline
\hline
\multicolumn{11}{c}{\bf{Band 9 - 6}} \\ 
\hline
\hline
J2228-170725-B96-1deg&2017-07-25&05:59:58&J2253+1608&3.26&-0.718&33&0.27&19.3&15.8&2943\\
J2228-170725-B96-6deg&2017-07-25&06:46:10&J2253+1608&3.26&-0.718&33&0.27&16.1&13.2&2851\\
J0449-170725-B96-5deg$^{c}$&2017-07-25&11:06:01&J0522$-$3627&5.18&-0.184&22&0.43&17.2&14.1&2339\\
J0449-170725-B96-7deg$^{c}$&2017-07-25&11:52:14&J0522$-$3627&5.18&-0.184&22&0.45&28.7&23.5&2340\\
J0449-170725-B96-12deg$^{c}$&2017-07-25&13:41:00&J0522$-$3627&5.18&-0.184&26&0.45&18.9&15.4&2887\\
J2228-170825-B96-3deg&2017-08-25&04:35:49&J2253+1608&3.13&-0.705&46&0.41&34.0&27.8&4461\\
J2228-170828-B96-6deg$^e$&2017-08-28&07:06:40&J2253+1608&3.13&-0.705&47&0.47&35.4&28.9&4792\\
\hline

\end{tabular}
}
\begin{tablenotes}
Notes: The tests are ordered into band pairs, where the first band is that of the target and the following that of the calibrator for the B2B blocks only. Tests are identified by the target, the observing date (YYMMDD), the B2B frequency pair and the related in-band calibrator separation angle in the naming scheme. Baseline length is the maximal projected value and are rounded to the nearest meter. The flux and spectral index ($\alpha$) of the DGC source and the phase RMS given in degrees relate to the target observing frequency. The expected phase RMS is that measured on the DGC over $\sim$30\,s and combines all baselines $>$1.5\,km. An uncertainty of the order 10-20\,\% is not unreasonable. \\
$^a$5 antennas flagged, $^b$Uses DGC block from J2228$-$170926-B73-1deg, $^c$Only has/uses one DGC block, $^d$Only has one in-band block, $^e$6 antennas flagged, $^f$All DGC blocks failed, used one from J0633$-$170930-B73-1deg, $^g$Only has one B2B block, $^h$One in-band block flagged as source $>$85$^{\circ}$ elevation, $^i$Last DGC block flagged.
\end{tablenotes}
\label{tab0}
\end{center}
\end{table*}

\begin{table}[ht!]
\begin{center}
  \caption{Central frequencies for the SPWs.}
  {\footnotesize
\begin{tabular}{@{}lcc@{}}
\hline
SPW  & High Frequency (GHz)  &  Low Frequency (GHz)   \\ 
  &   &  B2B (calibrator only)   \\ 

\hline
\hline
\multicolumn{3}{c}{\bf{Band 7 (B2B pair Band3)}} \\  
\hline
\hline
0  & 278.013   &  81.083 \\
1  & 279.971   & 90.048\\
2  & 290.013   & 100.830\\
3  & 291.971   & 102.041\\
\hline
\hline
\multicolumn{3}{c}{\bf{Band 8 (B2B pair Band4) }} \\ 
\hline
\hline
0  & 393.026   & 126.486\\
1  & 394.984   & 128.444\\
2  & 405.026   & 138.486\\
3  & 406.984   & 140.444\\
\hline
\hline
\multicolumn{3}{c}{\bf{Band 9 (B2B pair Band4) }} \\ 
\hline
0  & 678.744   & 144.070\\
1  & 680.702  & 146.028\\
2  & 678.744   & 154.101\\
3  & 680.702   & 156.101\\
\hline
\hline
\multicolumn{3}{c}{\bf{Band 9 (B2B pair Band6) }} \\ 
\hline
0  & 681.728   & 216.118\\
1 & 683.686   & 218.076\\
2  & 681.728   & 232.118\\
3  & 683.686   & 234.076\\
\hline
\hline

\end{tabular}
}
\label{tab1}
\end{center}
\end{table}

\subsection{Differential Gain Calibration}
\label{dgc_sec}
DGC is the only way to determine the instrumental phase difference between the high frequency and low frequency bands. It is accomplished by observing a bright quasar while switching quickly between the two bands. DGC is detailed more thoroughly in \citet{Asaki2020} and \citet{Asaki2020b}, although we provide a short overview for context. The phase difference between the DGC source observed at two frequencies is a delay term dominated by atmospheric variations and an instrumental offset. Because of the changing atmosphere, fast switching between frequencies is required such that a transfer of the low frequency phase solutions can be used to calibrate the high frequency atmospheric variations that one assumes have not changed significantly over the switching time. Under this assumption, the remaining delay is considered to be entirely instrumental and is thus the characteristic band-offset, hereafter the DGC solution. It is imperative that the DGC solution is not contaminated by any atmospheric phase variations which could be detrimental to B2B phase calibration. We explore this in more detail in Section \ref{dgcdirect}.

In these tests we use fast frequency switching with a 24\,s cycle time where $\sim$9\,s is spent on-source at the high- and low-frequencies respectively, with around 3\,s to switch between frequencies. All DGC source fluxes at the high frequency are $>$2.5\,Jy in all bands. The separations between the DGC sources and science targets are $<$30$^{\circ}$ in all cases. The DGC source is only used to find the instrumental offset, akin to how a bright source within some tens of degrees of a target is used to solve for the instrumental bandpass response in general operations, and thus the separation is not critical \citep{Asaki2020}. Two DGC solutions are found for the start and end DGC blocks and applied using a linear interpolation to correct for any slow drifts \citep{Asaki2020b}.

\subsection{Data Reduction and Processing}
With a forward look to commissioning and implementation of B2B, a standardized script was developed based on the ALMA quality assessment procedures using {\sc casa} \citep{McMullin2007}. Each of the tests was conducted in the same sequence such that the script automatically identified all required SPWs and scans to use for each source at each frequency and proceed in a step wise manner. Although the process of calibration was therefore almost entirely automated, it did not preclude checking the data and solutions during the reduction steps.

Automatic flags are typically generated in standard ALMA science observations when any system based issues occur. However, due to the nature of these tests with fast cycle times and frequency switching (in the B2B case), some of the typical online flags were not stored, and thus manual checking and flagging of all datasets were undertaken before calibration, imaging and analysis. The data reduction follows a relatively similar process to standard ALMA data reduction except for added stages to address the frequency switching for the B2B datasets. 

\subsubsection{In-band reduction}
In short, WVR, system-noise temperature (Tsys) and antenna position corrections are applied followed by the aforementioned manually set flags. The DGC source high-frequency scans are used to calibrate the bandpass response and also to provide a single flux amplitude calibration, i.e. there is no secondary temporal amplitude scaling bootstrapped from the phase calibrator gains, as to provide a fair comparison with the B2B blocks. Normal phase interpolation is conducted as the phase calibrator and target source are observed at the same frequency (i.e. standard phase referencing). Note that all high frequency SPWs are combined when obtaining the phase calibrator solution to maximize the signal-to-noise ratio (S/N). Ideally each SPW used for in-band calibration should generally be calibrated individually because any combination of SPWs averages the residual delays, which are then considered as phase solutions at an average frequency. Phase solution error (in radians) follow 1/(S/N) and are non-negligible for low S/N values, and so in some observations the combination is absolutely required because the phase calibrator was partially flagged, or was weaker than expected (e.g. at band 9). 

\subsubsection{B2B reduction}
For B2B calibration, the Tsys and WVR corrections are applied to both the high- and low-frequencies, while antenna position corrections and flags are applied as per in-band reduction. The DGC source is used first to calibrate the low- and high-frequency bandpass response and to provide a single flux amplitude calibration only at the high-frequency. No flux calibration is required at the low-frequency as we use only the phase solutions. Subsequently we solve for the DGC solution as outlined above. The only remaining correction to make now is that for the atmospheric fluctuations. Here, phase calibration transfers the phase solutions from the low frequency phase calibrator to the high frequency target scaled by the multiplicative ratio between frequencies, $\nu_h/\nu_l$ (similar to FPT, where $\nu_h$ and $\nu_l$ are the high- and low-frequencies respectively). For the harmonic-switching setup employed, the ratio is an integer value, although it is not a requirement in general for ALMA using normal frequency switching. The interpolation and scaling {\bf entirely} is handled by the `linearPD' interpolation option in the {\sc applycal} task within {\sc casa}, that fully solves for any phase ambiguities before scaling (G. Moellenbrock - private communication). We note that the calibrator low-frequency SPWs are averaged together in all tests. 

\subsubsection{Self-calibration}
All target sources are point source quasars and so we also perform self-calibration on the raw data. Again, we combine all SPWs in this process to boost the S/N. Self-calibration is undertaken in order to directly compare an ideal calibration, free of any residual phase errors, against both the in-band and B2B phase referencing techniques. Generally for self-calibration we use the integration timescale of $\sim$1\,s, as most of the targets are bright enough. However, for a number of observations we are limited to using the scan timescale $\sim$9\,s to achieve a sufficient S/N as the quasar targets become fainter above band 7 \citep{Cornwell1981,Brogan2018}. The slight caveat is that the very short timescale atmospheric fluctuations cannot be corrected out, although if the phase RMS is low (over a 9\,s scan) then the coherence loss is negligible. The band 9 targets cannot be self-calibrated as the S/N is too low, and thus we estimate the flux from the band 7 and 8 observations as part of this study. We also perform self-calibration on the DGC sources. 


\subsection{Measuring the Phase stability}
\label{ssfmeas}
Our observational strategy groups in-band and B2B blocks together as part of one observation sequence to ensure the direct comparability of the in-band and B2B phase calibration techniques as they are taken under the same observing conditions. However, to allow the comparison between different observations taken on different days and under different conditions we must consider the phase stability. 

The spatial-structure-function (SSF) $D_{\phi}(b)$ is the dispersion of the atmospheric phase as a function of baseline length $b$ measured over the entire time interval of a given observation ($t_{\rm obs}$, see also \citealt{Wright1996,Carilli1999,Matsushita2017}) and is related to the phase RMS by $\sigma_{\phi}$=$\sqrt{D_{\phi}(b)}$. Here we use the same metric but impose a specific timescale average: 

\begin{equation}
  \label{eqn2}
\sigma_{\rm{\phi}}(b,t) = \sqrt{D_{\phi}(b,t)} = \langle (\phi(x+b) - \phi(x))^2 \rangle_{t} ^{1/2} (deg),
\end{equation}

\noindent where $(\phi(x+b) - \phi(x))$ is the atmospheric phase difference between the two antennas, and the angle brackets represent an ensemble average over the timescale $t$, our chosen specific time range for the averaging period ($t$=$t_{\rm obs}$ results in the classic SSF). Crucially, we make all phase RMS assessments after application of the WVR solutions. This is because the WVR solutions are applied on the integration timescale and correct the phase fluctuations for all sources.

We refer to our first phase RMS measure we refer to as the expected phase RMS. This is established using a time interval, $t$, close to the cycle-time ($t_{\rm cyc}$) but on non-phase referenced data. This provides a representative value of the phase variations that will likely remain in the data after ideal phase referencing, because phase referencing only corrects fluctuations longer than the cycle time, and hence variations $\lesssim t_{\rm cyc}$ remain largely unchanged. We calculate the expected phase RMS using only the low frequency scans of the DGC source as these provide the highest S/N. Because our observing sequence has fixed and repetitive scan lengths, if we exclude the high frequency scan in-between two consecutive low-frequency scans we measure the expected phase RMS\footnote{For these phase RMS calculations we must account for a non-zero degrees mean phase as we use only WVR corrected data and must exclude phase offsets and only include the short-term phase variations over the selected scans. We therefore use a standard deviation calculation for each of the DGC scan pairs rather than a true RMS statistic.} over $\sim$30\,s, which is the closest match to the phase referencing cycle time of $\sim$24\,s. The final expected phase RMS is the average value from all paired low frequency scans, but scaled to the high-frequency at which the target it observed. This is also used to derive the expected coherence via Equation \ref{eqn1}. The expected phase RMS values are indicated in Table \ref{tab0}.


Second, we measure the residual phase RMS for the DGC source. The residual phase RMS is measured after B2B phase referencing has occurred, using the full time duration of the DGC source observations (block 1 and 6 combined, $t$=$t_{\rm obs}$), and only the high frequency data. The residual phase RMS provides the true value of phase fluctuations remaining in the data. We detail the effect of the DGC source phase residuals in Section \ref{dgcdirect}.

Finally, we measure the residual phase RMS of the target sources pre- and post-calibration for the respective in-band and B2B target data. These calculations also use the entire duration of the observations on each target (blocks 2 and 4 for in-band, and blocks 3 and 5 for B2B data). The pre- and post-calibration values, combined with the expected phase RMS allow us to investigate how effective phase referencing is and contrast that with what was expected. This is discussed further in Section \ref{disc}.

\subsection{Imaging}  
All of the targets are imaged in an automatic fashion using the {\sc clean} command in {\sc casa}. The beam pixel sizes are chosen to be five times smaller than the synthesized beam (band 9 uses seven time smaller) and we use square maps with sizes of 512$\times$512 or 1024$\times$1024 pixels, dependent on resolution - higher resolution images use more pixels. Briggs {\citep{Briggs1995} robust weighting 0.5 is used for all images, as the optimal balance of resolution and sensitivity, and this is the current default for ALMA imaging in quality assessment reduction. {\sc clean}ing is undertaken within a 15 pixel radius circular region in the center of the map using a fixed number of 50 iterations, which is sufficient for the central point sources. The peak flux density and integrated flux are measured within the same circular aperture to parameterize the source, while the map noise is taken within an annulus between radii of 15 pixels out to 250 or 500 pixels depending on the image size 512 or 1024 respectively. The sources are also fitted with a 2D Gaussian in the image plane within the central region of the map. All self-calibrated images are measured in the same manner as those made with phase referencing. 

\begin{table*}[ht!]
\begin{center}
  \caption{Parameters of the in-band - B2B observations where the same nearby phase calibrator was used for both phase referencing techniques.}
  {\footnotesize
\begin{tabular}{@{}lllrrrrrrr@{}}
\hline
Name & Target & Calibrator & Sep. & \multicolumn{2}{c}{Peak (mJy/beam)} & \multicolumn{2}{c}{Flux (mJy)} & \multicolumn{2}{c}{Noise (mJy/beam)} \\ 
      &              &                  & (deg)       & In-Band & B2B & In-Band & B2B & InBand & B2B \\
\hline
\hline
\multicolumn{10}{c}{\bf{Band 7 - 3}} \\   
\hline
\hline
J2228-170827-B73-1deg&J2228$-$0753&J2229$-$0832&0.68&48.17&51.36&49.40&52.68&0.08&0.11\\
J2228-170829-B73-1deg&J2228$-$0753&J2229$-$0832&0.68&49.98&49.03&50.80&50.27&0.12&0.13\\
J0449-170829-B73-2deg&J0449$-$4350&J0440$-$4333&1.67&85.51&88.61&90.32&94.01&0.26&0.27\\
J2228-170830-B73-1deg&J2228$-$0753&J2229$-$0832&0.68&51.29&49.67&52.35&50.94&0.13&0.15\\
J2228-170917-B73-1deg&J2228$-$0753&J2229$-$0832&0.68&41.41&40.80&43.81&42.69&0.15&0.15\\
J0633-170917-B73-1deg$^a$&J0633$-$2223&J0634$-$2335&1.25&84.65&73.35&109.39&104.85&0.48&0.63\\
J2228-170926-B73-1deg&J2228$-$0753&J2229$-$0832&0.68&43.06&43.63&45.11&45.14&0.11&0.12\\
J0449-170928-B73-2deg&J0449$-$4350&J0440$-$4333&1.67&90.88&96.71&100.72&103.75&0.38&0.41\\
J0633-170930-B73-1deg$^a$&J0633$-$2223&J0634$-$2335&1.25&151.11&139.62&156.23&150.05&0.35&0.56\\
\hline
\hline
\multicolumn{10}{c}{\bf{Band 8 - 4}} \\ 
\hline
\hline
J1709-170717-B84-1deg&J1709$-$3525&J1713$-$3418&1.37&47.61&42.75&69.08&68.20&0.67&0.78\\
J1259-170717-B84-1deg&J1259$-$2310&J1258$-$2219&0.85&190.86&178.32&196.61&187.19&0.66&0.93\\ 
J0633-170718-B84-1deg$^a$&J0633$-$2223&J0634$-$2335&1.25&152.83&149.71&157.13&149.30&0.42&0.63\\ 
J2228-170819-B84-1deg$^a$&J2228$-$0753&J2229$-$0832&0.68&37.96&33.63&40.25&35.95&0.41&0.43\\
J2228-170820-B84-1deg&J2228$-$0753&J2229$-$0832&0.68&19.88&21.02&28.92&24.25&0.48&0.53\\
\hline
\hline
\multicolumn{10}{c}{\bf{Band 9 - 4}} \\
\hline
\hline
J2228-170717-B94-1deg&J2228$-$0753&J2229$-$0832&0.68&13.51&13.10&16.67&10.62&2.99&3.01\\ 
\hline
\hline
\multicolumn{10}{c}{\bf{Band 9 - 6}} \\ 
\hline
\hline
J2228-170725-B96-1deg&J2228$-$0753&J2229$-$0832&0.68&16.78&9.89&15.14&11.66&1.87&1.68\\
\hline

\end{tabular}
}
\begin{tablenotes}
Notes: The peak, integrated flux, and noise levels are indicated for the target source after calibration. Band 9 flux accuracy limited by thermal noise.\\
$^a$B2B nosier than in-band by $>$30\,\%.
\end{tablenotes}
\label{tab2}
\end{center}
\end{table*}

\subsection{Image Assessment Criteria: image coherence, fidelity, dynamic range, defect}
\label{assmt}
We define four measures to assess the results: image coherence, fidelity, dynamic range, and defect. The concept of coherence was introduced in Section \ref{intro} as the estimated coherence of the visibility data calculable based on the expected phase RMS of the data. Here we define the image coherence which is measured directly by comparing the peak flux densities of the target from phase referenced images using in-band and B2B calibration with the respective self-calibrated images  \citep{Asaki2020b}. In a number of VLBI studies this is often referred to as the fractional (peak) flux recovered, or peak-ratio \citep[e.g.][]{Dodson2009,Rioja2011b,Rioja2015,MartiVidal2010}. To provide a measure of image fidelity, the accuracy of the reconstructed sky brightness distribution, we simply use the measure of peak flux density divided by the integrated flux. Given the targets are point sources these should be equal for an ideal calibration (peak flux density in Jy/beam and integrated flux in Jy). Any deviation from this equality indicates a spreading of the flux in the image and a poorer representation of the point source target. The dynamic range is simply the image peak divided by the image noise level. Finally, we judge whether there are significant defects that shift the source central position or alter the structure by using the Gaussian fits. Strictly speaking these parameters are somewhat correlatated, in that poor phase calibration will result in a low image coherence and will cause image defects leading to a low image fidelity and thus overall poor image quality.

\section{Results and Comparions}
\label{res}
The following subsections address goals (1) and (2) outlined at the end of Section \ref{intro}. Where relevant we introduce the assessment criteria and measures of the expected and residual phase RMS to make comparisons. We also divide the observations by maximal baseline length, separated into short-baseline - $<$3.7\,km, mid-baseline - 3.7 to 8.5\,km, and long-baseline - $>$8.5\,km groups. This roughly divides into the C43$-$7, C43$-$8 and C43$-$9/10 operational ALMA array configurations\footnote{https://almascience.eso.org/documents-and-tools/cycle5/alma-technical-handbook/view}. Predominantly, the longer baseline observations, $>$5\,km, are made in band 7 as the PWV conditions during the available testing time were unsuitable for good transmission for higher bands.

\subsection{Comparison of in-band and B2B calibration with the same calibrator}
\label{direct}
There are a total of 16 observations divided into different frequency pairings, B7-3 (9), B8-4 (5), B9-4 (1) and B9-6 (1). Table \ref{tab2} lists these datasets along with the peak flux densities and integrated flux values and image noise. The calibrators used in these observations can be considered as almost ideal as they are extremely close to the target, they were selected to be $\sim$1$^{\circ}$ away. The maximal target-to-calibrator separation is 1.67$^{\circ}$, while nine of the datasets target J2228$-$0753 for which the phase calibrator, J2229$-$0832, is separated by only 0.68$^{\circ}$. 

In Figure \ref{fig2} the comparisons of the image peak flux density (left), integrated flux (center) and the image map noise (right) of the targets calibrated with the B2B and in-band techniques from each dataset are presented. Note each point shows parameters of the same target calibrated by both B2B and in-band techniques within one observation (the in-band value on the x-axis while the B2B value is on the y-axis). The peak and integrated fluxes comparing in-band with B2B images are on average equal to within 5\,\% for band 7 and  7\,\% for band 8 (blue and purple symbols). The in-band images typically have higher values. This is as one might expect considering the extra DGC step for the B2B technique, that could introduce minor phase uncertainties (see Section \ref{dgcdirect}). Band 9 image parameters are consistent given the uncertainties and low number statistics. Most noticeable are the discrepancies in the image map noise where in five cases the B2B noise is more than 25\,\% larger than the corresponding in-band observation (see Table \ref{tab3}). The worse case shows a $\sim$60\,\% increase for J0633-170930-B73-1deg. This observation has a failed in-band block and DGC block due to a hardware instability in the telescope system, although the B2B data did record without obvious errors. Such an extreme outlier does not appear to be the norm. We note that when using only one DGC block any linear change of the DGC solution cannot be corrected. We also cannot rule out effects of system instabilities. After self-calibration the target image noise values are more consistent between the in-band and B2B data. 

Figure \ref{fig3} shows the comparisons of the image assessment criteria (Section \ref{assmt}), the image coherence (left), fidelity (center) and the dynamic range (right). We find that the B2B image coherence values typically are within 6\,\% (bands 7 and 8) of the in-band values, which again are generally better. For fidelity the B2B images are within 3\,\% (band 7) and 8\,\% (band 8) of the in-band values. Still this is indicative of the B2B technique closely matching that of the standard in-band phase referencing calibration. The dynamic range panel mirrors the image noise panel from Figure \ref{fig2}, where B2B images with worse noise are those with a lower dynamic range. Table \ref{tab3} presents all the plotted parameters. None of the fitted target positions show discrepancies larger than one-third of their respective synthesized beams for the band 7 and band 8 data. Good fits are not achieved for band 9 as they are limited by thermal noise.

\subsubsection{Determining the effect of DGC}
\label{dgcdirect}
The only critical difference between in-band and B2B phase referencing is the extra step of DGC to correct the instrumental band-offset. We can surmise that any inaccuracies in the DGC solution are responsible for discrepancies when comparing in-band and B2B images. Because the instrumental offset can only be found by DGC we have no ideal value to compare with. However, we can investigate whether uncorrected atmospheric variations negatively impact the DGC solution. Figure \ref{fig4} plots the difference between the in-band and B2B image coherence as a percentage against the residual phase RMS of the DGC high-frequency data after B2B phase referencing. There are nine band 7 and 8 datasets with higher in-band image coherence values (excluding J0633-170930-B73-1deg which has significantly higher B2B image noise). The black dashed line is a linear fit to the logarithm of the residual phase RMS against the coherence difference of those datasets. The trend implies that any remaining atmospheric fluctuations during DGC have a negative impact on the DGC solution, and thus marginally degrade the final B2B target source images. The degradation follows: $17.9\, \log_{10}(\sigma_{\rm{\phi\,DGC}})\,-19.4$, and is small, only $\sim$4\,\% and $\sim$7\,\% when the residual phase RMS $\sigma_{\rm{\phi\,DGC}}$= 20 and 30$^{\circ}$ respectively. At a residual phase RMS $\sigma_{\rm{\phi\,DGC}}$= 12$^{\circ}$ the coherence difference approaches zero. Except the aforementioned outlier J0633-170930-B73-1deg there are no datasets with a residual phase RMS $<$10$^{\circ}$, at which level the expected coherence is $>$98\,\%. One could surmise that in-band and B2B images are equal to within an image coherence of 1-2\,\% at such low residual phase RMS levels, and any differences likely relate to uncertainties in the phase solutions in the calibration irrespective of the technique used. Provided that the residual phase RMS for the DGC source is minimized, using fast frequency switching, then the DGC stage should not be significantly detrimental to the final target image quality for B2B calibration (also see Section \ref{conds})

\begin{table*}[ht!]
\begin{center}
  \caption{Image coherence, Image Fidelity, Dynamic range, DGC source image coherence and Expected coherence parameters of the in-band - B2B observations where the same nearby phase calibrator was used for phase referencing.}
  {\footnotesize
\begin{tabular}{@{}lllrrrrrrr@{}}
\hline
Name  & \multicolumn{2}{c}{Image Coherence} & \multicolumn{2}{c}{Fidelity} & \multicolumn{2}{c}{Dyn. Range} & DGC image & Expected  \\ 
            & In-band & B2B & In-band & B2B & In-band & B2B & Coherence & Coherence \\
            \hline
\hline
\multicolumn{9}{c}{\bf{Band 7 - 3}} \\   
\hline
\hline
J2228-170827-B73-1deg&0.97&0.97&0.97&0.97&571.97&465.74&0.98&0.98\\
J2228-170829-B73-1deg&0.94&0.95&0.98&0.98&408.21&390.42&0.98&0.97\\  
J0449-170829-B73-2deg&0.90&0.90&0.95&0.94&325.34&323.53&0.98&0.97\\
J2228-170830-B73-1deg&0.95&0.93&0.98&0.98&381.92&330.98&0.96&0.96\\
J2228-170917-B73-1deg&0.92&0.90&0.95&0.96&272.38&267.35&0.94&0.94\\
J0633-170917-B73-1deg&0.60&0.52&0.77&0.70&176.73&117.04&0.61&0.67\\
J2228-170926-B73-1deg&0.95&0.96&0.95&0.97&403.18&352.20&0.98&0.98\\
J0449-170928-B73-2deg&0.84&0.90&0.90&0.93&239.66&234.61&0.97&0.96\\
J0633-170930-B73-1deg&0.97&0.92&0.97&0.93&427.81&250.09&0.99&0.99\\
\hline
\hline
\multicolumn{9}{c}{\bf{Band 8 - 4}} \\ 
\hline
\hline
J1709-170717-B84-1deg&0.72$^a$&0.64$^a$&0.69&0.63&71.30&54.93&0.81&0.89\\
J1259-170717-B84-1deg&0.90&0.85&0.97&0.95&289.26&192.23&0.92&0.91\\
J0633-170718-B84-1deg&0.93&0.89&0.97&1.00&360.00&236.27&0.94&0.96\\
J2228-170819-B84-1deg&0.90$^a$&0.85$^a$&0.94&0.94&92.74&78.78&0.93&0.95\\
J2228-170820-B84-1deg&0.63$^a$&0.63$^a$&0.69&0.87&41.82&39.77&0.75&0.76\\
\hline
\hline
\multicolumn{9}{c}{\bf{Band 9 - 4}} \\
\hline
\hline
J2228-170717-B94-1deg&0.73$^b$&0.71$^b$&0.81&1.23&4.52&4.35&0.75&0.85\\
\hline
\hline
\multicolumn{9}{c}{\bf{Band 9 - 6}} \\ 
\hline
\hline
J2228-170725-B96-1deg&0.91$^b$&0.53$^b$&1.11&0.85&8.99&5.88&0.94&0.96\\
\hline
\end{tabular}
}
\begin{tablenotes}
Notes: Fidelity for the band 9 data is above 1.0 as the weak detection is thermal noise dominated.\\
$^a$Indicates that the image coherence was calculated against the self-calibrated image that used the scan length solution interval (9\,s) to achieve a self-calibration signal-to-noise $\gtrsim$3.\\ $^b$ Indicates that the image coherence was calculated against the expected band 9 source flux after extrapolation from the self-calibrated band 7 and band 8 images.
\end{tablenotes}
\label{tab3}
\end{center}
\end{table*}

\subsubsection{Close calibrators for phase referencing}
\label{closedirect}
In an attempt to isolate and investigate the effect of the small calibrator-to-target separation angles ($<$1.67$^{\circ}$ for these close calibrator data), we compare the target image coherence with the DGC source image coherence, and these with the respective expected coherence values (Figure \ref{fig5}). Here the DGC source acts as the ideal phase referencing case because there is no position change for the phase-referencing, only the temporal phase transfer using B2B. In calculating the expected coherence, we make a correction to the expected phase RMS measured on the DGC sources to account for the different elevation of the target sources, $\sigma_{\rm{\phi}}(\theta_{\rm tar})$ = $\sigma_{\rm{\phi}}(\theta_{\rm dgc})\,sin(\theta_{\rm dgc})/sin(\theta_{\rm tar})$, where $\theta_{\rm tar}$ and $\theta_{\rm dgc}$ are the elevations of the target and DGC sources respectively \citep{Holdaway1997, Butler1997}. The maximum elevation difference between the DGC sources and the targets is 29$^{\circ}$, although over half the datasets have elevation differences $<$10$^{\circ}$. The elevation corrections generally change the expected phase RMS by $<$10\,\%, well within the assigned 20\,\% uncertainties of the expected RMS phase calculation itself. The target image coherence is the average value from both B2B and in-band images which are nearly equal. 

The left panel of Figure \ref{fig5} shows that the target image coherence values are comparable with the DGC source image coherence values, on average within 5\,\% and 11\,\% for bands 7 and 8 respectively. Separated into mid- and long-baselines for band 7, the differences are 4\,\% and 6\,\% on average. As one might expect the DGC image coherence values are marginally better, and thus the small reduction in target image coherence is as a result of the small target to phase calibrator separation angle. Roughly one or two percent could be be attributed to detrimental effect of DGC on the B2B images as the target image coherence used is the average of in-band and B2B ones. The center and right panels of Figure \ref{fig5} compare the target and DGC source image coherence parameter with the expected coherence calculated using the expected phase RMS, respectively. These coherence measures are very similar and on average for the target images are within 6\,\% excluding images differing by $>$20\,\%. The band 9 are consistent with these findings considering the larger uncertainties. The DGC source image coherence and expected coherence values are almost equal and are presented in the last two columns of Table \ref{tab3}.  

\begin{figure*}  
\begin{center}
  \includegraphics[width=18.5cm]{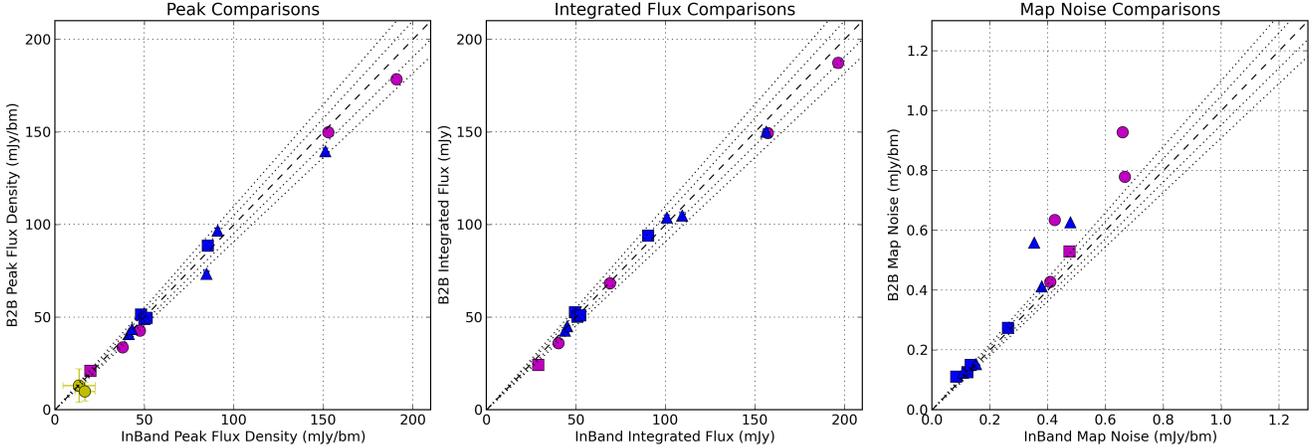}  
\caption{Comparisons of the image peak flux densities, integrated fluxes and map noise from the in-band and B2B calibrated datasets that use the same phase calibrator. The color relates to the target frequency, blue is band 7, purple is band 8 and yellow is band 9. The symbols are representative of the maximal baseline lengths, circles are for maximal baselines between 2.0\,km and 3.7\,km, squares are for 3.7\,km to 8.5\,km, while triangles are for baselines $>$8.5\,km. The dashed line is that of equality, whereas the dotted lines either side denote discrepancies of $\pm$5 and 10\,\%. In terms of peak flux density and integrated flux most datasets are within a 6\,\%, in-band values being marginally larger. There are a few notable outliers in terms of map noise, where B2B calibration results in worse values. Uncertainties in peak flux density and integrated flux are plotted at three times the image map noise.}
\label{fig2} 
\end{center}
\end{figure*}

The fundamental result here is that independent of phase calibration technique, in-band or B2B phase referencing, the target images have comparable parameters and that the DGC stage does not have a significant detrimental effect for B2B observations. The comparability of the target images with the DGC source images indicates that small phase calibrator separation angles ($<1.67^{\circ}$) can recover to within 7\,\% (5\,\% - excluding the two images with differences $>$15\,\% at band 8) of the maximal coherence possible for bands 7 and 8 combined. Furthermore, provided that the phase RMS is measured over a timescale similar to the proposed cycle time, using a strong point source, one can already establish a representative value for expected phase RMS achievable after phase referencing. This {\bf in turn} translates to an expected image coherence, which can be evaluated before the observations have even taken place, of course under the premise that a close by calibrator would be used. Figure \ref{fig6} presents images of the target source J2228$-$0753 in band 7 (top), band 8 (middle) and band 9 (bottom) using the B2B (left) and in-band (right) phase referencing techniques that all used the same phase calibrator, J2229$-$0832, separated by only 0.68$^{\circ}$.

\begin{figure*}  
\begin{center}
  \includegraphics[width=18.5cm]{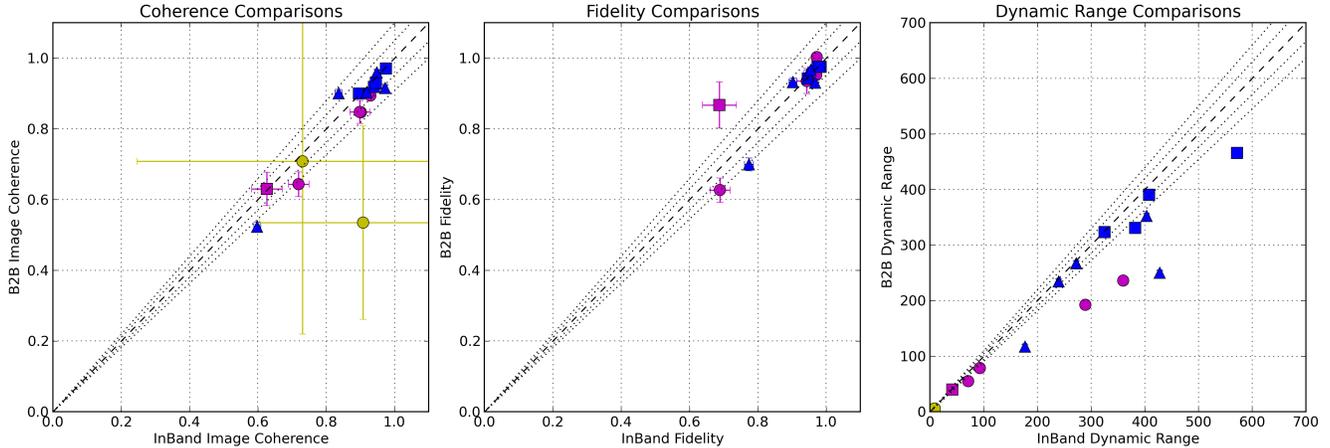} 
\caption{Comparisons of the image coherence (as a fraction, 1.0 = 100\,\%), fidelity and dynamic range from the B2B and in-band calibrated data that use the same phase calibrator. Analogously to Figure \ref{fig2} image coherence and fidelity values are in good agreement, indicating B2B calibration can result in images as good as in-band calibration. The dynamic range is worse for a few B2B datasets, caused by the underlying increase in map noise. Colors (bands) and symbols (maximal baseline length) are the same as Figure \ref{fig2}. Uncertainties are propagated from those in the peak flux density and integrated flux (three times the image map noise). Band 9 are not plotted in the central panel with values $>$1.} 
\label{fig3} 
\end{center}
\end{figure*}

\begin{figure}  
\begin{center}
\includegraphics[width=8.5cm]{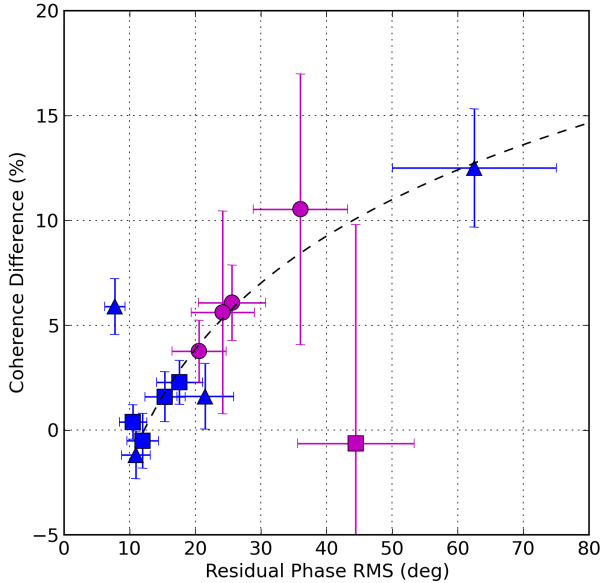} 
\caption{Difference in image coherence (as a percentage) for in-band and B2B calibrated data using the same close calibrators as a function of residual phase RMS of the DGC source after B2B phase referencing. Colors (bands) and symbols (maximal baseline length) are the same as Figure \ref{fig2}. Positive values indicate a better in-band image coherence while negative values indicate that the B2B image coherence is higher. Band 9 are not included considering the large uncertainties in coherence. Uncertainties are propagated from those of the peak flux density (three times the image map noise). There is a trend of increasing coherence difference as a function of residual phase RMS. The black dashed line is a linear fit to the logarithm of the residual phase against the positive coherence difference values. J0633-170930-B73-1deg, where the B2B data was notably nosier, is excluded from the fit - triangle at x=7.7$^{\circ}$. If the residual phase RMS can be minimized then DGC has only a small effect on B2B calibrated images. However, if observing conditions are marginal and the phase RMS remained high, in-band calibration would be preferable provided that a close calibrator was available.}
\label{fig4} 
\end{center}
\end{figure}

\begin{figure*}  
\begin{center}
  \includegraphics[width=18.5cm]{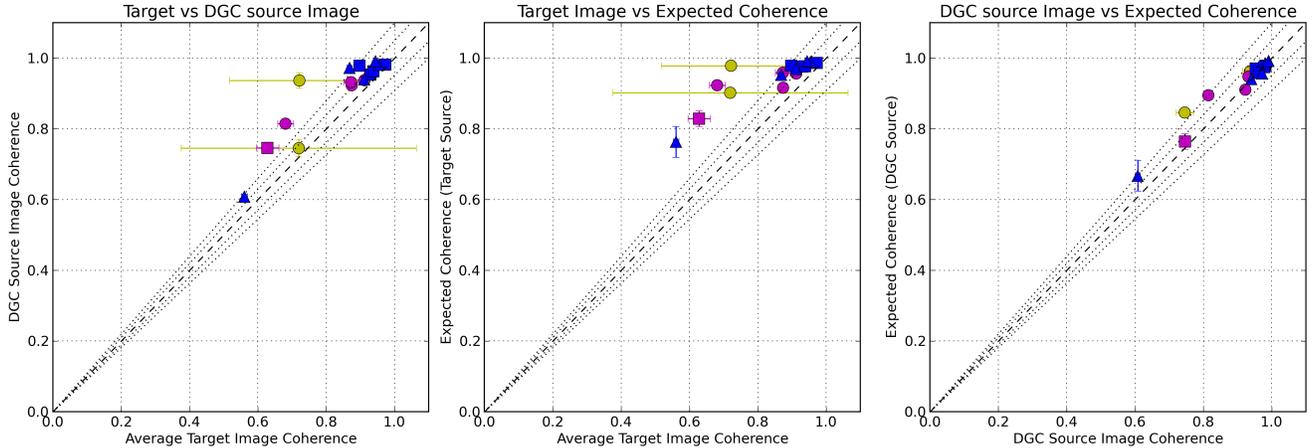} 
\caption{Comparisons of the average target image coherence with the DGC source image coherence (left) and expected coherence (center), and (right) the DGC source image coherence with the expected coherence. Coherence values are presented as a fraction, 1.0 = 100\,\%. The DGC source image coherence represents the best value achievable using phase referencing but without any source position change, while the expected value is calculated from the expected phase RMS of the DGC source measured over $\sim$30\,s, via Equation \ref{eqn1}. In the center panel the expected phase RMS is scaled to account for elevation difference for the target source before computing the expected coherence. The similarity between the target and DGC source image coherence suggests that phase referencing with small calibrator separation angles generally has little detrimental effect on target images. The right panel indicates that the coherence calculated from the expected phase RMS matches well with that found in the images of the DGC sources. Colors (bands) and symbols (maximal baseline length) are the same as Figure \ref{fig2}. Note band 9 data target image coherence values are calculated against the expected band 9 flux extrapolated from the band 7 and band 8 self calibrated data. Uncertainties for the image coherence values are propagated from those in the peak flux density of three times the image map noise, while that of the expected coherence is from a 20\,\% variability in the expected phase RMS.}
\label{fig5} 
\end{center}
\end{figure*}

\begin{figure*}  
\begin{center}
  \includegraphics[width=18.5cm]{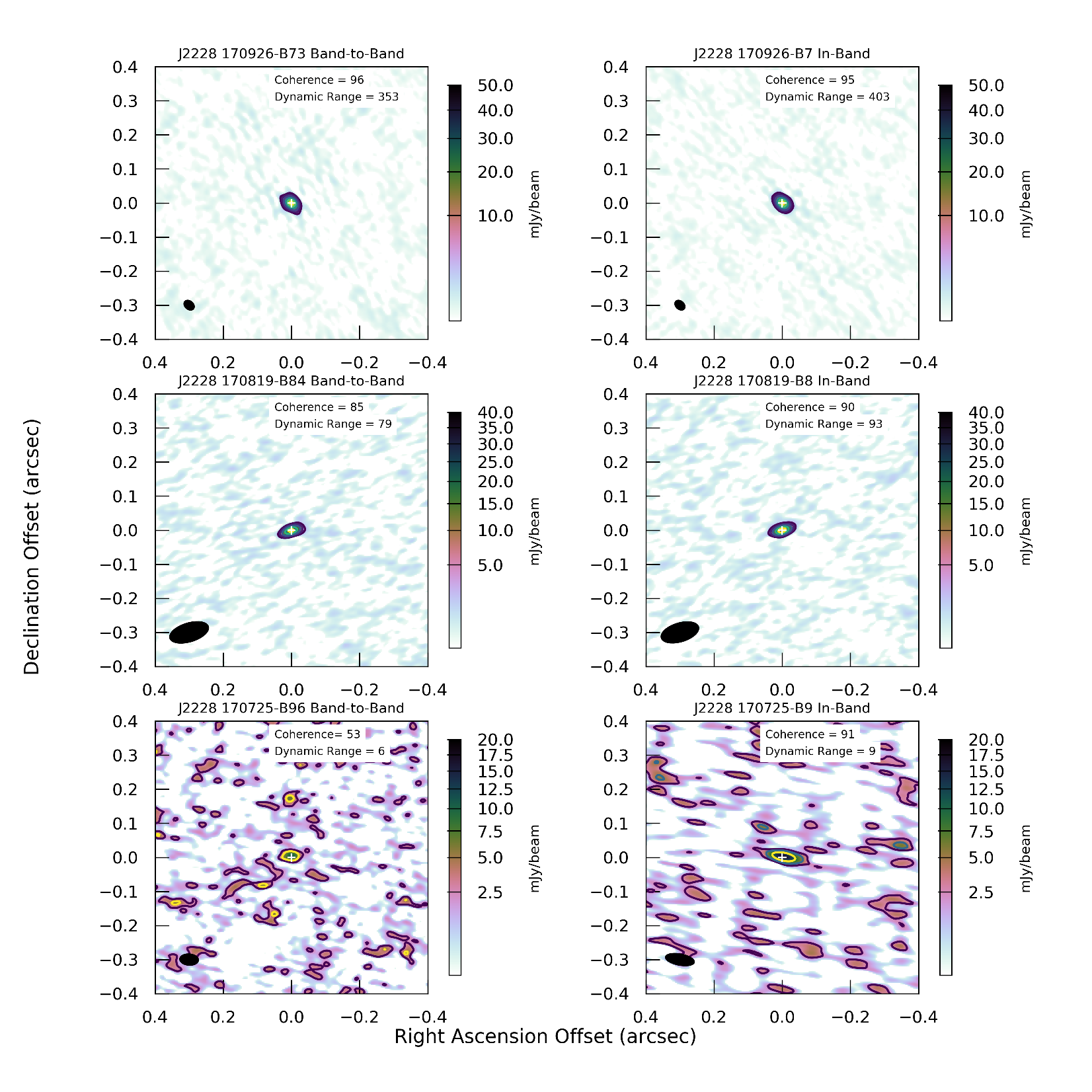}  
\caption{Images of the target source J2228$-$0753 at band 7 (top), band 8 (middle) and band 9 (bottom) using the B2B (left) and in-band (right) techniques using the same phase calibrator J2229$-$0832, separated by only 0.68$^{\circ}$ in all cases. All observations have an expected phase RMS $<$30$^{\circ}$ and should all achieve a coherence $>$87\,\%. The coherence (\%) and dynamic range values are listed to the top right of each panel. Note the band 9 images are only weak detections and that the in-band beam is larger than that for B2B as some automatic flags occurred due to the weak phase calibrator. Overall the B2B and in-band images are comparable. We note the band 9 B2B is marginally worse but only used one DGC block. The color ranges are scaled to 50, 40 and 20\,mJy for panels top to bottom, while the contour levels are at 2.5, 5, 10, 20, 30, 40 and 50\,mJy/beam in all plots. }
\label{fig6} 
\end{center}
\end{figure*}

\subsection{Investigation of phase calibrator separation angles}
\label{calsep}
In the following sub-sections we present a number of results in order to investigate the effect of phase calibrator separation angle building upon the results presented in Section \ref{direct}. 

\subsubsection{Comparison of In-band using distant calibrators vs. B2B using close calibrators}
\label{conds}
The underlying philosophy of the B2B technique is that more optimal calibrator choices are available. Using a lower frequency provides a higher chance of phase calibrators being stronger and closer to a given target source \citep[see][]{Asaki2020}. The remaining test observations were arranged such that the B2B phase calibrator is separated at most by 1.67$^{\circ}$ from the target, whereas the in-band calibrator can range from 2.42$^{\circ}$ to 11.65$^{\circ}$ away. In each observation the target source in the B2B and in-band blocks is the same but a different calibrator is used. Table \ref{tab4} lists these observations along with the peak flux density, integrated flux values and image noise. For the B2B blocks there are B7-3 (12), B8-4 (10) and B9-6 (6) frequency pairs.

Following the same analysis as Section \ref{direct}, in Figure \ref{fig7} (left) we now find that the peak flux density values of the B2B calibrated images noticeably exceed those using in-band calibration (except one band 9 dataset). Building from the fact that when using the same phase calibrator the in-band images were marginally better than B2B images, here the complete opposite trend occurs. The central panel shows that the integrated fluxes are generally comparable between the B2B and in-band images, while the right panel mirrors the left panel, in that the majority of in-band images have higher image map noise values compared to the B2B images. There are five in-band observations with an image noise $>$30\,\% higher than the B2B ones. The reduction of peak flux density and the increase in noise for the in-band images, while still somewhat recovering a similar integrated flux as the B2B images, suggests a loss of coherence due to poorer phase correction. Even though B2B images can suffer from a few percent degradation due to the DGC solution (see Section \ref{dgcdirect}), the effect of using distance calibrators is much more detrimental. Figure \ref{fig8} provides a visual representation of this by showing the images of J2228-0753 taken on 2017 September 26 in band 7 using the longest baselines. The left panel shows the B2B images that always use a calibrator only 0.68$^{\circ}$ from the target, while the in-band images on the right, top to bottom, use calibrators 0.68, 3.04, and 6.02$^{\circ}$ away respectively (see also Figure 10 in \citealt{Asaki2020}). The B2B images remain largely unchanged during the three observations, while the in-band images gradually degrade, the poor phase correction evidently blurs the target images \citep[][see also]{Carilli1999}.

\begin{table*}[ht!]
\begin{center}
  \caption{Parameters of the B2B and in-band observations where a more distant phase calibrator was used for the in-band phase referencing blocks compared to the B2B blocks.}
{\scriptsize
\begin{tabular}{@{}lllllrrrrrrr@{}}
\hline
Name & Target & B2B & Sep. & In-band & Sep. & \multicolumn{2}{c}{Peak (mJy/beam)} & \multicolumn{2}{c}{Flux (mJy)} & \multicolumn{2}{c}{Noise (mJy/beam)} \\ 
      &              &  Calibrator   & (deg)    &  Calibrator   & (deg)     & In-band & B2B & In-band & B2B & In-band & B2B \\
\hline
\hline
\multicolumn{12}{c}{\bf{Band 7 - 3}} \\ 
\hline
\hline
J2228-170829-B73-3deg&J2228$-$0753&J2229$-$0832&0.68&J2225$-$0457&3.04&41.10&45.13&47.43&45.40&0.16&0.14\\ 
J2228-170829-B73-6deg&J2228$-$0753&J2229$-$0832&0.68&J2246$-$1206&6.02&41.09&48.07&49.34&48.09&0.15&0.13\\ 
J0449-170829-B73-3deg&J0449$-$4350&J0440$-$4333&1.67&J0451$-$4653&3.08&90.56&90.95&96.66&94.73&0.22&0.20\\
J2228-170830-B73-3deg&J2228$-$0753&J2229$-$0832&0.68&J2225$-$0457&3.04&47.55&50.63&51.52&51.10&0.18&0.14\\
J0449-170830-B73-5deg&J0449$-$4350&J0440$-$4333&1.67&J0515$-$4556&5.12&84.73&96.26&100.19&100.37&0.50&0.27\\
J0449-170830-B73-7deg&J0449$-$4350&J0440$-$4333&1.67&J0428$-$3756&7.08&82.57&93.99&87.83&95.72&0.33&0.17\\
J0633-170917-B73-4deg&J0633$-$2223&J0634$-$2335&1.25&J0620$-$2515&4.11&14.40&43.35&64.42&68.06&0.67&0.67\\
J2228-170926-B73-3deg&J2228$-$0753&J2229$-$0832&0.68&J2225$-$0457&3.04&37.27&42.41&44.77&46.04&0.17&0.14\\
J2228-170926-B73-6deg&J2228$-$0753&J2229$-$0832&0.68&J2246$-$1206&6.02&30.07&44.93&43.90&46.37&0.29&0.12\\
J0449-170929-B73-5deg&J0449$-$4350&J0440$-$4333&1.67&J0515$-$4556&5.12&30.79&49.59&61.88&59.00&0.64&0.53\\
J0633-170930-B73-4deg&J0633$-$2223&J0634$-$2335&1.25&J0620$-$2515&4.11&126.82&131.94&153.55&141.44&0.77&0.66\\
J0633-171001-B73-9deg&J0633$-$2223&J0634$-$2335&1.25&J0609$-$1542&8.72&4.59&45.66&15.56&56.56&0.50&0.58\\
\hline
\hline
\multicolumn{12}{c}{\bf{Band 8 - 4}} \\ 
\hline
\hline
J1709-170717-B84-2deg&J1709$-$3525&J1713$-$3418&1.37&J1717$-$3342&2.42&40.84&54.55&66.43&70.98&0.74&0.67\\
J1709-170717-B84-11deg&J1709$-$3525&J1713$-$3418&1.37&J1626$-$2951&10.65&34.39&45.87&66.87&68.71&0.79&0.67\\
J2228-170717-B84-7deg&J2228$-$0753&J2229$-$0832&0.68&J2236$-$1433&6.92&11.27&24.93&27.20&25.63&0.42&0.62\\
J1259-170717-B84-8deg&J1259$-$2310&J1258$-$2219&0.85&J1245$-$1616&7.57&139.04&171.42&178.97&178.22&1.20&0.84\\
J1259-170718-B84-11deg&J1259$-$2310&J1258$-$2219&0.85&J1316$-$3338&11.12&114.95&170.03&155.27&179.62&1.23&0.94\\
J0633-170718-B84-4deg&J0633$-$2223&J0634$-$2335&1.25&J0620$-$2515&4.11&91.60&141.69&104.58&144.45&0.82&0.88\\
J0633-170718-B84-9deg&J0633$-$2223&J0634$-$2335&1.25&J0609$-$1542&8.72&123.32&146.52&145.07&142.44&1.20&0.76\\
J0633-170718-B84-6deg&J0633$-$2223&J0634$-$2335&1.25&J0648$-$1744&5.84&138.08&153.24&148.74&153.72&0.67&0.78\\
J2228-170820-B84-3deg&J2228$-$0753&J2229$-$0832&0.68&J2225$-$0457&3.04&15.38&25.09&27.34&21.12&0.53&0.60\\
J2228-170820-B84-10deg&J2228$-$0753&J2229$-$0832&0.68&J2158$-$1501&10.37&6.43&28.06&21.70&30.95&0.51&0.48\\
\hline
\hline
\multicolumn{12}{c}{\bf{Band 9 - 6}} \\ 
\hline
\hline
J2228-170725-B96-6deg&J2228$-$0753&J2229$-$0832&0.68&J2246$-$1206&6.02&9.76&13.11&14.76&15.40&1.97&1.92\\
J0449-170725-B96-5deg&J0449$-$4350&J0440$-$4333&1.67&J0515$-$4556&5.12&66.82&37.45&77.74&54.71&4.05&4.44\\ 
J0449-170725-B96-7deg&J0449$-$4350&J0440$-$4333&1.67&J0428$-$3756&7.08&37.97&67.04&76.30&84.41&4.04&4.29\\ 
J0449-170725-B96-12deg&J0449$-$4350&J0440$-$4333&1.67&J0403$-$3605&11.65&19.65&47.88&13.68&46.41&4.77&3.87\\ 
J2228-170825-B96-3deg&J2228$-$0753&J2229$-$0832&0.68&J2225$-$0457&3.04&7.57&14.36&15.13&13.03&1.48&1.62\\
J2228-170828-B96-6deg$^a$&J2228$-$0753&J2229$-$0832&0.68&J2246$-$1206&6.02&-&16.57&-&11.09&3.94&3.45\\ 
\hline
\end{tabular}
}
\begin{tablenotes}
Notes: The calibrator names and separation angles are listed for both the B2B and in-band blocks. The peak flux density, integrated flux, and image noise levels are indicated for the target source after phase referencing calibration using both techniques. There are no B9-4 datasets as the sources were too weak to be imaged. \\
$^a$ The target could not be imaged with in-band calibration.
\end{tablenotes}
\label{tab4}
\end{center}
\end{table*}

Figure \ref{fig9} presents the image coherence, image fidelity and dynamic range comparisons for the B2B images using a close calibrator against the in-band images using a more distant calibrator. The negative effect of using more distant phase calibrators for the in-band observations is clear, where in almost all cases the B2B images have better image coherence and image fidelity than the comparison in-band images. Interestingly, a number of the lower image coherence datasets ($<$60\,\%) show that the B2B values are often a factor of two better than the comparative in-band values. This suggests that poorer atmospheric stability also decorrelates the calibrators scans potentially introducing phase errors, that when combined with larger phase calibrator separation angles looking through a different line-of-sight has a compound effect on the image quality.

In Figure \ref{fig10} we plot the image coherence of the in-band (left) and B2B (center) observations separately against the DGC source image coherence. The left panel highlights that the in-band data, with more distant phase calibrators, have notably reduced image coherence values compared to that achievable in an ideal case. The majority of in-band image coherence values are over 15\,\% worse than the DGC source image coherence. The B2B images (center panel) using the close calibrators have image coherence values predominantly within $\sim$10\,\% of the DGC image coherence, as already show for close calibrators in Figure \ref{fig5} (left). The right panel shows the DGC source image coherence compared with the expected coherence, again, indicating that the estimated coherence is a good proxy for that expected in an ideal phase referencing case. Table \ref{tab5} lists the parameters as plotted.

Finally, in Figure \ref{fig11} we plot the quality parameters of image position uncertainty and excess source size with respect to the synthesized beam each dataset. The left and right panel pairs separate the in-band and B2B datasets. The spread of centroid fit positions is more variable for the in-band calibrated data which use more distant phase calibrators. A few in-band images exceed a discrepancy of more than 1/3 of the synthesized beam (dashed circle). Two of these are long-baseline band 7 observations. In contrast, none of the B2B images exceed this limit. The discrepancies in fitted source size excess are also seen for the in-band images, many of which have measurable blurring or spreading of the emission $>$15\,\%, while some surpass 1/3 of the synthesized beam size. In general the B2B image excess sizes are $<$10\,\%. We note that J0633-170917-B73-4deg is the only dataset with a size excess $>$1/3 of the beam for the B2B data, although it was taken in very unstable phase conditions. The corresponding in-band image size excess is $>$1 beam.

\begin{figure*}  
\begin{center}
  \includegraphics[width=18.5cm]{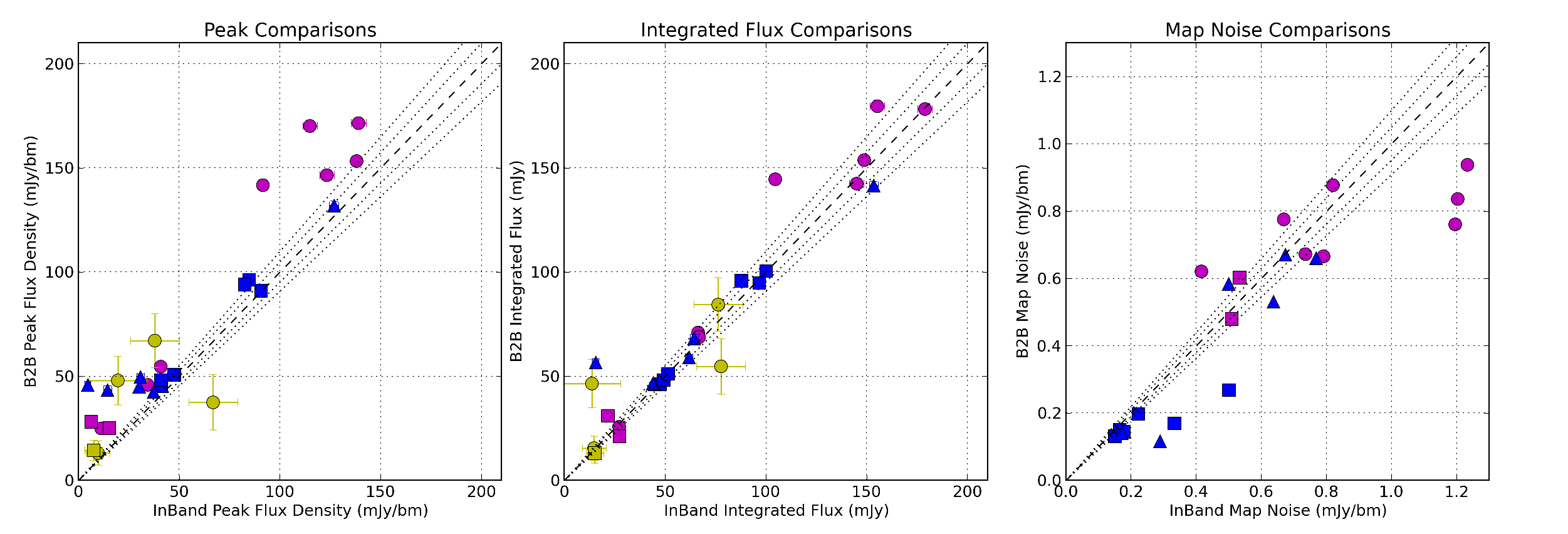} 
\caption{Comparisons of the peak flux densities, integrated fluxes and map noise from the in-band and B2B calibrated datasets that use different phase calibrators. The B2B calibrators are always within 1.67$^{\circ}$ while the in-band ones range from 2.42 to 11.65$^{\circ}$. Most strikingly, compared to Figure \ref{fig2} almost all the B2B image values are better than the in-band ones, i.e. have a higher peak flux density. This clearly indicates a worse calibration for the in-band block due to using a more distant phase calibrator, while the B2B blocks use a close one, i.e. B2B with close calibrators will result in better images. For band 9 images the noise values are off the scale. Color (bands) and symbols (baseline length) are the same as Figure \ref{fig2}. The dashed line is that of equality, whereas the dotted lines either side denote discrepancies of 5 and 10\,\%. Uncertainties in peak flux density and flux are plotted as three times the image noise.}
\label{fig7} 
\end{center}
\end{figure*}

\begin{figure*}  
\begin{center}
  \includegraphics[width=18.5cm]{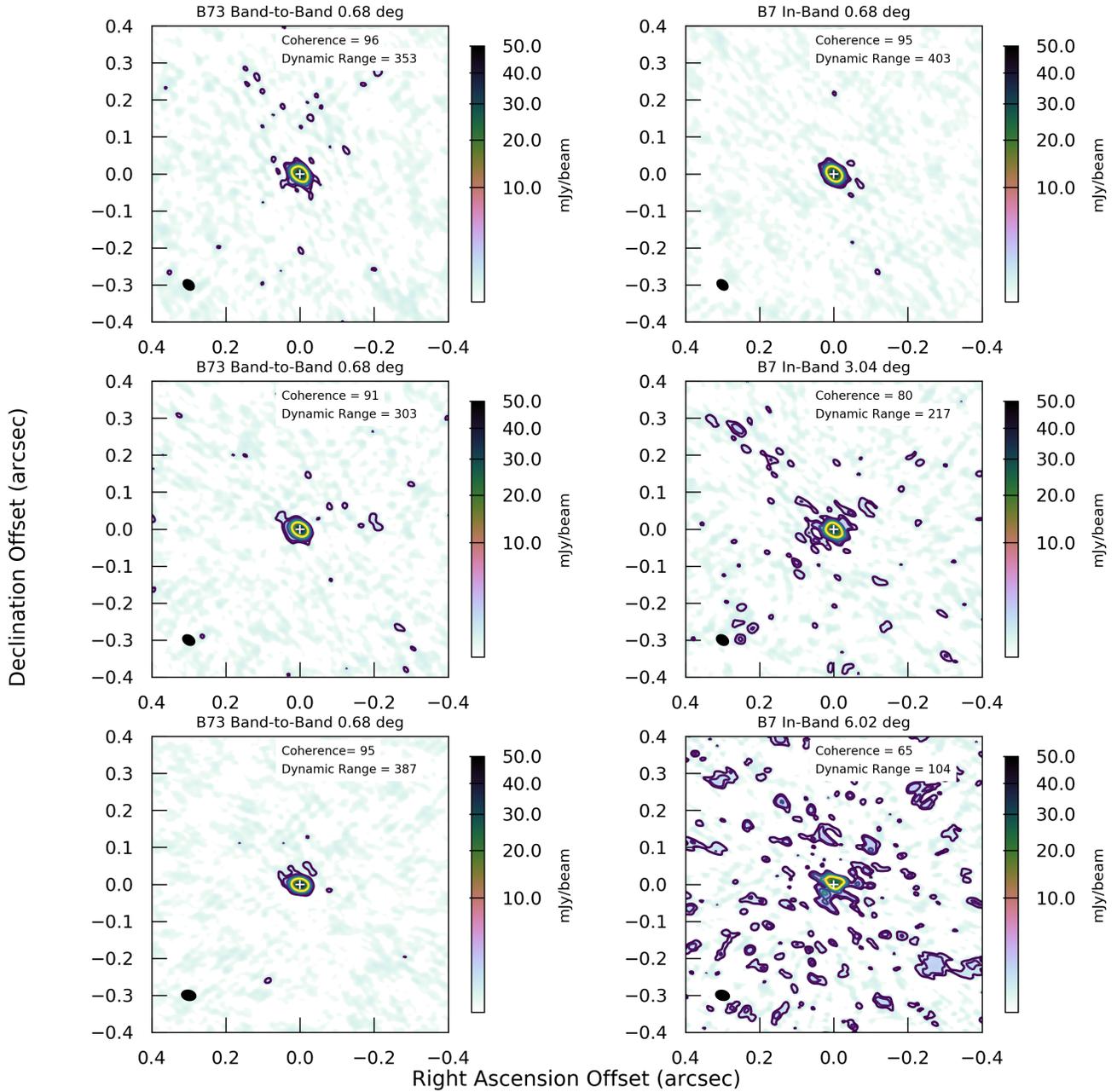} 
\caption{Images of the target source J2228-0753 in band 7 using the longest baselines taken on 2017 September 26 in three consecutive observations, top: B2B and in-band have the same calibrator at 0.68$^{\circ}$ separation, middle: B2B keeps the 0.68$^{\circ}$ calibrator but in-band uses a more distant calibrator at 3.04$^{\circ}$, bottom: B2B uses the 0.68$^{\circ}$ calibrator while in-band now uses a distant calibrator at 6.02$^{\circ}$. The color ranges are all scaled to 50\,mJy, while the contour levels are at 0.5, 1.0, 2.5, 5.0 and 10.0\,mJy/beam in all plots to highlight the weakest emission and an image defects. Note 0.5\,mJy/beam corresponds to $\sim$3-4\,$\sigma$ for all images except the in-band 6.02$^{\circ}$ image which has a higher noise value of 0.29\,mJy/beam. The B2B images with the same calibrator remain largely unchanged from top to bottom in the three observations while the in-band images gradually degrade in terms of both image coherence and dynamic range as a result of poor phase correction with increasing calibrator-to-target separation angle. The beam sizes are of the order 35$\times$25\,mas.}
\label{fig8} 
\end{center}
\end{figure*}

\begin{figure*}  
\begin{center}
  \includegraphics[width=18.5cm]{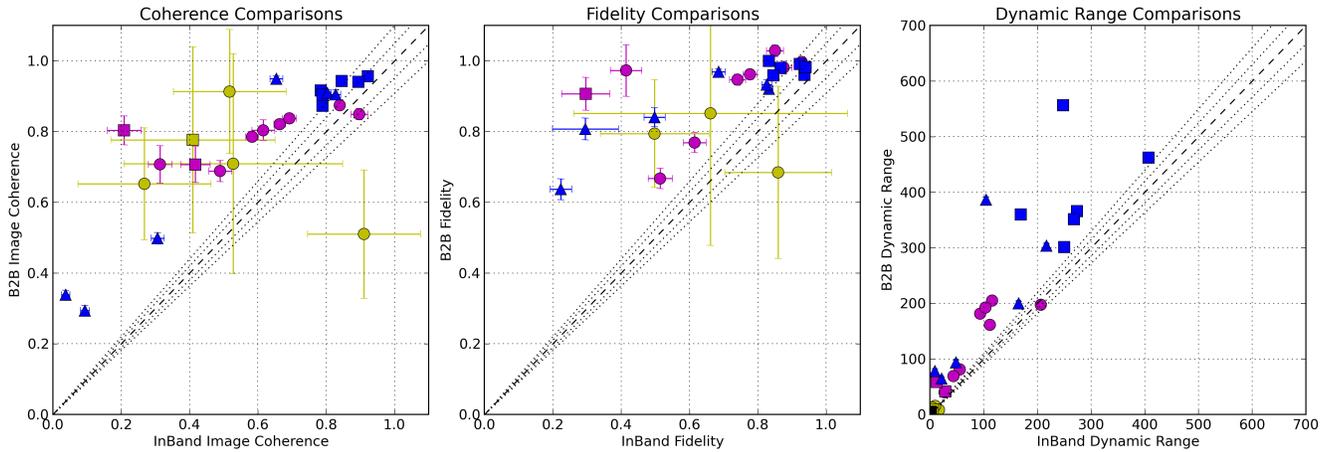} 
\caption{Comparisons of the image coherence (as a fraction), fidelity and dynamic range from the in-band and B2B calibrated data that use different phase calibrators. The image coherence and fidelity values indicate that the B2B calibration is superior. The dynamic range is worse for in-band datasets as caused by the underlying larger map noise values. Colors (bands) and symbols (maximal baseline length) are as Figure \ref{fig2}. We note that some datasets have fidelity values $>$1 due to negatives regions summed within the aperture when establishing the flux. Uncertainties are propagated from those in the peak flux density and integrated flux (three times the image map noise).}
\label{fig9} 
\end{center}
\end{figure*}

\begin{figure*}  
\begin{center}
  \includegraphics[width=18.5cm]{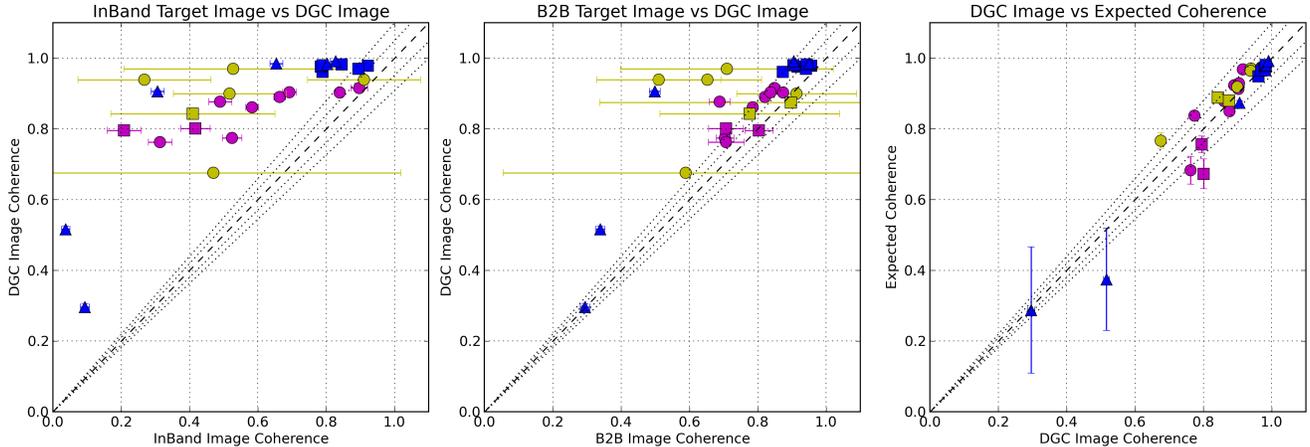}  
\caption{Comparisons of the in-band (left) and B2B (center) target image coherence with the DGC source image coherence, and the DGC source image coherence against the expected coherence (right). Coherence values are plotted as fractions. The DGC source image coherence represents the ideal case using phase referencing but without any source position change, while the expected value is calculated from the expected phase RMS. The in-band image coherence values are always notably less than the DGC source values - as a result of using distant calibrators. Typically the B2B observations nearly achieve the ideal image coherence values - i.e. phase referencing with small calibrator separation angles generally has little detrimental effect on target images. Colors (bands) and symbols (maximal baseline length) are as Figure \ref{fig2}. The band 9 data target image coherence values are calculated against the expected band 9 flux extrapolated from the band 7 and band 8 self calibrated data. Uncertainties in image coherence are propagated from those in the peak flux density and integrated flux (three times the image map noise), while that for the expected coherences is related to a 20\,\% variability of the expected phase RMS.}
\label{fig10} 
\end{center}
\end{figure*}

\begin{figure*}  
\begin{center}
  \includegraphics[width=18.cm]{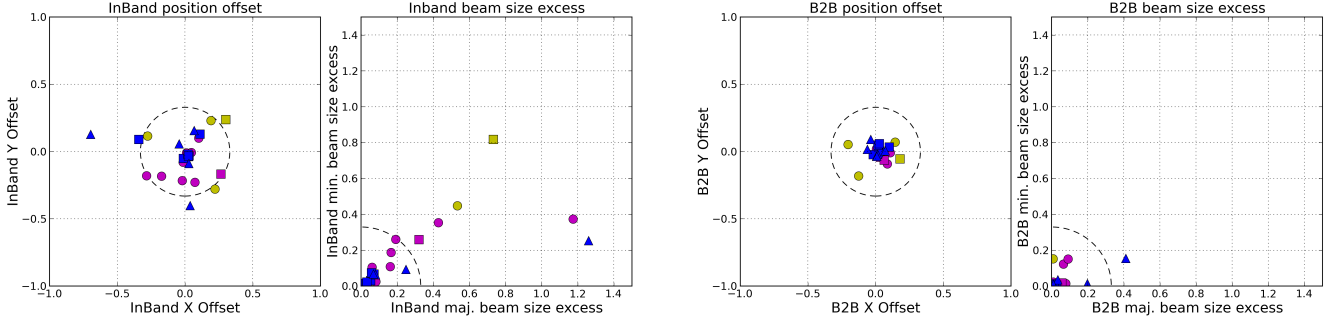}  
\caption{Image fit offsets and source size excess as a fraction of the image synthesized beam for in-band images (left) and B2B images (right). The in-band images calibrated using more distant phase calibrators have more noticeable offsets in the centroid source position compared to the B2B data calibrated with close calibrators. The in-band image fits also indicate a number of target source sizes are $>$10-15\,\% larger than the synthesized beam (i.e. blurred images) and some exceeded 1/3 of the beam. The dashed circle and arc indicate a measure of 1/3 of the synthesized beam. Note J0633-171001-B37-9deg is off the scale as the source image is not point-like and hence the fitted parameters are erroneous.}
\label{fig11}  
\end{center}
\end{figure*}

\begin{table*}[ht!]
\begin{center}
  \caption{Image coherence, Image Fidelity, Dynamic range, DGC source image coherence and Expected coherence parameters of the B2B - in-band observations where different phase calibrators were used.}
  {\footnotesize
\begin{tabular}{@{}lllrrrrrrr@{}}
\hline
Name  & \multicolumn{2}{c}{Image Coherence} & \multicolumn{2}{c}{Fidelity} & \multicolumn{2}{c}{Dyn. Range} & DGC image & Expected  \\
            & In-band & B2B & In-band & B2B & In-band & B2B & Coherence & Coherence \\

\hline
\hline
\multicolumn{9}{c}{\bf{Band 7 - 3}} \\   
\hline
\hline
J2228-170829-B73-3deg&0.79&0.87&0.87&0.99&249.94&332.24&0.95&0.95\\ 
J2228-170829-B73-6deg&0.78&0.90&0.83&1.00&273.57&365.91&0.96&0.97\\ 
J0449-170829-B73-3deg&0.92&0.96&0.94&0.96&406.85&462.27&0.98&0.97\\
J2228-170830-B73-3deg&0.89&0.94&0.92&0.99&268.12&351.27&0.97&0.96\\
J0449-170830-B73-5deg&0.79&0.90&0.85&0.96&169.24&359.87&0.98&0.97\\
J0449-170830-B73-7deg&0.85&0.94&0.94&0.98&247.86&556.51&0.98&0.98\\
J0633-170917-B73-4deg&0.09&0.29&0.22&0.64&21.36&64.68&0.30&0.29\\
J2228-170926-B73-3deg&0.80&0.91&0.83&0.92&216.58&303.5&0.98&0.98\\
J2228-170926-B73-6deg&0.65&0.95&0.69&0.97&104.15&386.68&0.98&0.98\\
J0449-170929-B73-5deg&0.31&0.50&0.50&0.84&48.32&93.53&0.91&0.87\\
J0633-170930-B73-4deg&0.83&0.90&0.83&0.93&164.96&199.97&0.99&0.99\\
J0633-171001-B73-9deg&0.04&0.34&0.30&0.81&9.20&78.22&0.52&0.37\\
\hline
\hline
\multicolumn{9}{c}{\bf{Band 8 - 4}} \\ 
\hline
\hline
J1709-170717-B84-2deg&0.52$^a$&0.70$^a$&0.61&0.77&55.51&81.14&0.77&0.84\\
J1709-170717-B84-11deg&0.49$^a$&0.69$^a$&0.51&0.67&43.46&68.95&0.88&0.85\\
J2228-170717-B84-7deg&0.31$^a$&0.71$^a$&0.41&0.97&27.09&40.15&0.76&0.68\\
J1259-170717-B84-8deg&0.66&0.82&0.78&0.96&115.58&204.96&0.89&0.92\\
J1259-170718-B84-11deg&0.58&0.79&0.74&0.95&93.21&181.28&0.86&0.88\\
J0633-170718-B84-4deg&0.9&0.85&0.88&0.98&111.61&161.57&0.92&0.97\\
J0633-170718-B84-9deg&0.69&0.84&0.85&1.03&103.14&192.65&0.90&0.93\\
J0633-170718-B84-6deg&0.84&0.87&0.93&1.00&206.47&197.59&0.90&0.91\\
J2228-170820-B84-3deg&0.42$^a$&0.71$^a$&0.56&1.19&28.83&41.65&0.80&0.67\\
J2228-170820-B84-10deg&0.21$^a$&0.80$^a$&0.30&0.91&12.63&58.5&0.80&0.76\\
\hline
\hline
\multicolumn{9}{c}{\bf{Band 9 - 6}} \\ 
\hline
\hline
J2228-170725-B96-6deg&0.53$^b$&0.71$^b$&0.66&0.85&4.95&6.84&0.97&0.97\\
J0449-170725-B96-5deg&0.91$^b$&0.51$^b$&0.86&0.68&16.51&8.43&0.94&0.97\\
J0449-170725-B96-7deg&0.52$^b$&0.91$^b$&0.50&0.79&9.41&15.62&0.90&0.92\\
J0449-170725-B96-12deg&0.27$^b$&0.65$^b$&1.44&1.03&4.12&12.37&0.94&0.96\\
J2228-170825-B96-3deg&0.41$^b$&0.78$^b$&0.50&1.10&5.12&8.85&0.84&0.89\\
J2228-170828-B96-6deg$^c$&-&0.90$^b$&-&1.49&1.70&4.81&0.87&0.88\\
\hline
\end{tabular}
}
\begin{tablenotes}
$^a$Indicates that the image coherence was calculated against the self-calibrated image that used the scan length solution interval (9\,s).\\ $^b$ Indicates that the image coherence was calculated against the expected band 9 source flux after extrapolation from the self-calibrated band 7 and band 8 images.\\ $^c$ The target could not be imaged with in-band calibration. 
\end{tablenotes}
\label{tab5}
\end{center}
\end{table*}

\subsubsection{Calibrator separation angle dependence}
\label{phaseangle}
Figure \ref{fig12} presents a number of in-band image parameters as a function of calibrator separation angle. The larger symbol sizes highlight images where the expected coherence should be $>$87\,\% after calibration. The top-left panel shows the ratio of the in-band image peak flux densities to the paired B2B image peak values, which all use close calibrators and should be assumed as the best achievable images when phase referencing. The bottom-left panel shows the in-band image coherence, i.e. the ratio of the image peak flux compared with that from the self-calibrated ideal in-band image. Both left panels are similar, and point to a general trend of in-band image peak flux degradation with separation angle. Moreover, the in-band coherence is much lower than expected ($<$87\,\%) for a number of the low expected phase RMS datasets. The trend is perhaps most compelling for band 8 and band 9 datasets which probe a greater range of separation angles. The band 7 long-baseline points (blue-triangles) suggest a steeper underlying slope when compared with the relatively constant value for mid-baselines (blue-squares), although there are few datasets over a narrow separation angle range. We further investigate differences with baseline length and frequency in Section \ref{freqBL}.

The central-top panel shows the ratio of in-band image noise to that achieved in the B2B images. The lighter symbols indicate data where either in-band or B2B blocks were missing, which would cause a notable noise change. By-eye, the band 7 longer baselines (blue-triangles) appear to increase sharply, while shorter baseline band 8 data increase gradually with separation angle (purple-circles). In general, the noise is worse for in-band images compared to B2B images, although there is notable scatter and no clear ensemble trend with separation angle. The bottom-central panel shows the ratio of the in-band dynamic range to the B2B image dynamic range. A clear decreasing trend with increasing separation angle is apparent with all frequency bands closely clustered. The main driver of this trend is likely the peak flux density. The top-right panel shows the image fidelity. Given that the in-band and B2B image integrated fluxes are generally in agreement, the shallow trend is also likely driven by the change in image peak flux density. Overall, increasing calibrator-to-target separation angles will reduced the recovered image peak flux densities.

The bottom-right panel of Figure \ref{fig12} shows the magnitude of the image position offset ($x_{\rm off}^{2} + y_{\rm off}^{2})^{1/2}$ against calibrator separation angle. Here $x_{\rm off}$ and $y_{\rm off}$ are the respective $x$ and $y$ position offsets of the peak flux density in the image from the central position, expressed as a fraction of the synthesized beam. A number of in-band calibrated images have a central position offset exceeding 1/3 of the beam (dotted line). The majority of the data hint at a trend of increasing position offset with increasing separation angle. The dashed line shows the fit for the position offsets after excluding those $>$0.25. The gradient (0.024$\pm$0.004) indicates a positional offset defect of the order $\sim$1/40 of the beam per degree.

\begin{figure*}  
\begin{center}
  \includegraphics[width=17cm]{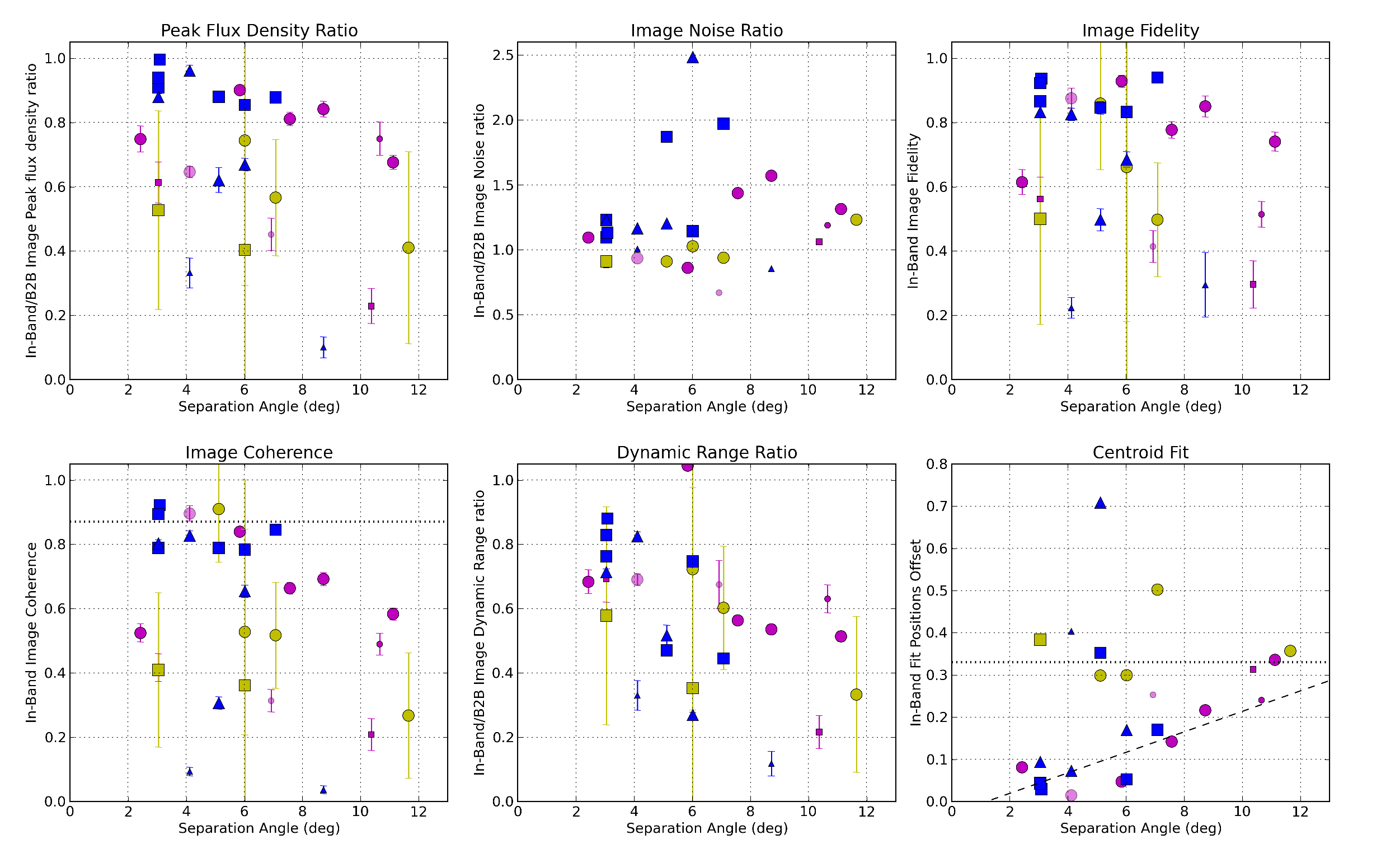}  
\caption{In-band image parameters as a function of calibrator to target separation angle. Colors (bands) and symbols (maximal baseline length) are the same as Figure \ref{fig2}. Top-to-bottom, left-to-right shows the ratio of the in-band peak flux density to the corresponding B2B target value, image coherence, the ratio of the in-band image map noise and ratio of the image dynamic range both to the corresponding B2B values respectively, the image fidelity, and the in-band image centroid position offset. The larger symbol sizes indicate datasets where the expected phase RMS should be $<$30$^{\circ}$, which translates an expected coherence 87\,\% (dotted line in bottom-left panel). In the top-middle panel the lighter symbols represent observations where either in-band or B2B blocks were entirely flagged. In the bottom-right panel the dotted line marks 1/3 of a beam, while the dashed line is the fit of position offset with separation angle. Excluding values $>$0.25, the gradient is 0.024. The uncertainties plotted are those that propagate from the error of three times the map noise for peak flux densities and integrated flux values. }
\label{fig12} 
\end{center}
\end{figure*}

\subsubsection{Dependence on Frequency and Baseline length}
\label{freqBL}

In Figure \ref{fig13} we plot the corrected in-band image coherence as a function of separation angle separated into the short- $<$3.7\,km (top), mid- 3.7 to 8.5\,km (middle), and long- $>$8.5\,km (bottom) baseline groups. Almost all the band 8 and band 9 observations are made with the shortest baselines, whereas the mid-baseline and long-baselines are dominated band 7 observations. We plot all datasets including those with the same phase calibrator. We  exclude data with only one in-band observing block from the fitting (see Table \ref{tab1}). The larger of the two symbol sizes again highlights those datasets that are expected to have low phase RMS ($<$30$^{\circ}$) if they were ideally calibrated. The corrected in-band image coherence values are the measured image coherence values (Tables \ref{tab5} and \ref{tab4}) corrected by the expected coherence loss (1.0 - expected coherence, where expected coherence is derived from the expected phase RMS measured on the DGC over $\sim$30\,s and scaled to the target elevation). For the low expected phase RMS data the corrections are 3\,\%, 7\,\% and 6\,\% on average for bands 7, 8 and 9 respectively. The underlying assumption is that phase referencing should have corrected the phase fluctuations down to the expected phase RMS level, and thus the images should all have achieved the expected coherence. Thereafter our correction should shift all datasets up to a corrected image coherence value of one. This however is not the case. We see clear trends of decreasing coherence with separation angle, meaning there are additional phase errors remaining, corrupting the image beyond the expected level and causing coherence losses.

We fit each frequency and baseline length independently using only the low expected phase RMS data. The fits are listed in Table \ref{tab6}. The y-intercept values are close to one, as we would expect. At zero separation there should be ideal phase transfer as there is no position change. Any remaining phase variations are due to the temporal phase-referencing which our correction to the coherence, based on phase RMS, accounts for. This ties with the previous findings that the DGC expected coherence and image coherence are almost equal when no position change is made during phase referencing (Section \ref{closedirect}, Figure \ref{fig5}-right). We note that the band 8 y-intercept is 0.92, skewed by two lower coherence datasets at separation angles of 1.3 and 2.4$^{\circ}$. Excluding these the y-intercept is increased to 0.99.  

A potentially interesting result is that the fitted gradients for band 9 short-baselines (-0.052$\pm$0.015) and band 7 long-baselines (-0.049$\pm$0.007) are consistent (within uncertainties), as are the band 8 short-baselines (-0.020$\pm$0.009) and band 7 mid-baselines (-0.024$\pm$0.007). The similarity suggests that these baselines and frequencies are similarly affected by increasing calibrator separation angle. The definitive results are that: image coherence degrades as a function of separation angle for data where calibration should have corrected the phase fluctuations to $<$30$^{\circ}$ phase RMS; the image coherence degradation is worse for longer baselines at the same observing frequency; the image coherence degradation is worse for higher frequencies using similar array configurations. We discuss these point further in Section \ref{disc}. If using these fits to estimate the final image coherence, any estimate must account for the coherence loss due to the expected phase RMS level achievable, i.e. the y-intercept would not be one: it should be the value of the expected coherence given an expected phase RMS (see Section \ref{reclim}). 

\begin{figure*}  
\begin{center}
  \includegraphics[width=18.5cm]{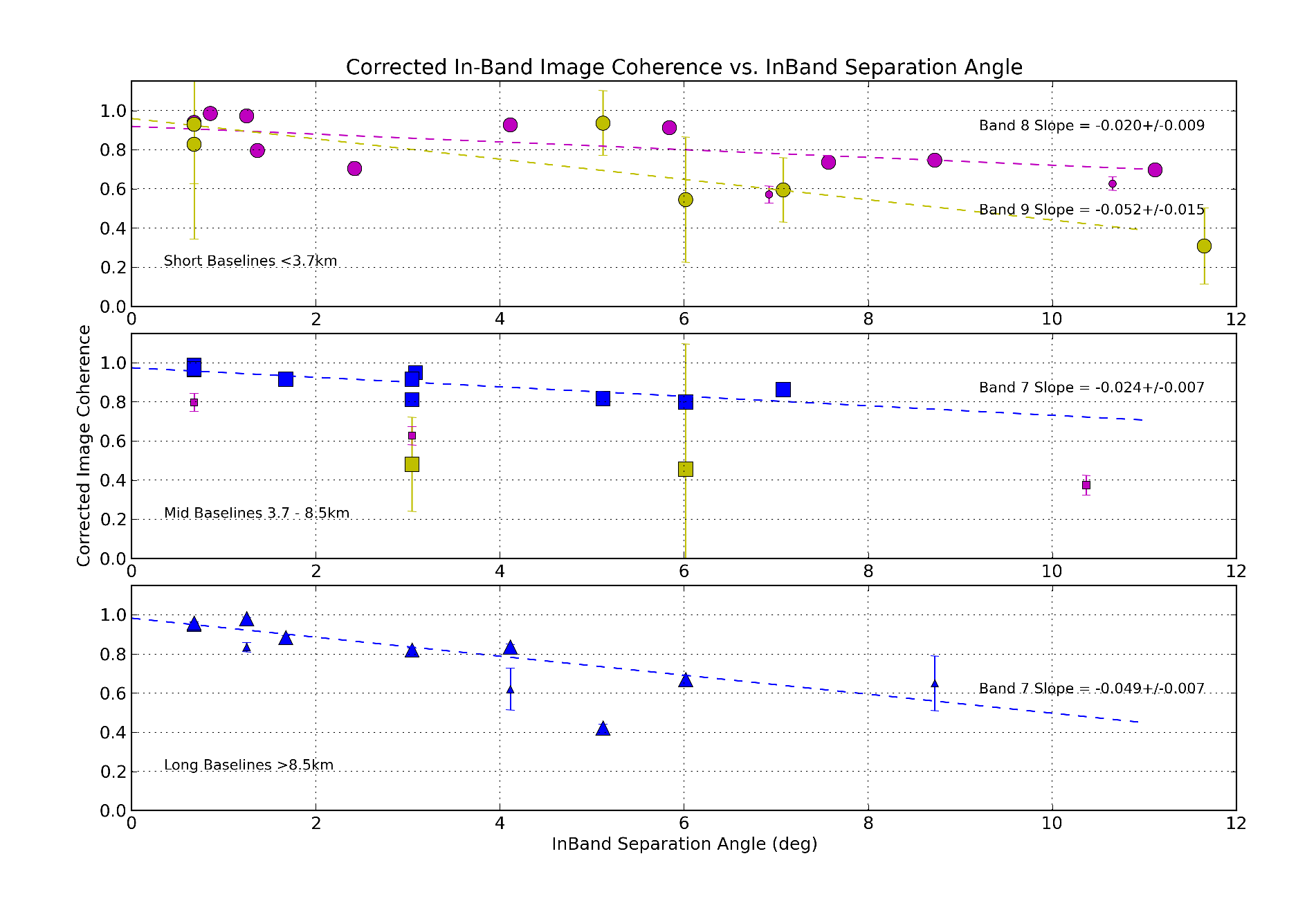}  
\caption{Corrected in-band image coherence (shown as a fraction, 1.0 = 100\,\%) as a function of in-band phase calibrator separation angle for short ($<$3.7\,km), mid (3.7$-$8.5\,km) and long ($>$8.5\,km) baseline length array configurations in the top, middle and bottom panels, respectively. The expected coherence loss is added to the observed image coherence to correct for the expected phase RMS, which should be the phase RMS remaining  after calibration. All datasets are plotted including those where the in-band and B2B blocks use the same phase calibrators (separation angles $<$1.67$^{\circ}$). These plots show a clear coherence degradation as a function of separation angle. The gradients, fitted individually for the different frequencies and baseline lengths are annotated on the panels. The larger of the two symbol sizes highlights those datasets where the expected phase RMS should be $<$30$^{\circ}$ and thus should have achieved a coherence of 87\,\%. The dashed lines are fits using only these low expected phase RMS data. Colors (bands) and symbols (maximal baseline length) follow Figure \ref{fig2}. Errors are propagated from: an uncertainty in the peak flux density that is three times the image noise; and a variation of 20\% in the phase rms. In some cases these are smaller than the symbol.}
\label{fig13} 
\end{center}
\end{figure*}

\begin{table}[ht!]
\begin{center}
  \caption{Linear fit parameters for the in-band corrected image coherence against in-band calibrator separation angle. }
  {\footnotesize
\begin{tabular}{@{}lcc@{}}
\hline
Baseline  length (km)   & Offset  &  Slope   \\ 
\hline
\hline
\multicolumn{3}{c}{\bf{Band 7}} \\  
\hline
\hline
3.7-8.5   & 0.975   & -0.024 $\pm$ 0.007\\ 
$>$8.5  &0.984  &  -0.049 $\pm$ 0.007\\
\hline
\hline
\multicolumn{3}{c}{\bf{Band 8 }} \\ 
\hline
\hline
$<$3.7 & 0.920   & -0.020 $\pm$ 0.009\\ 
\hline
\hline
\multicolumn{3}{c}{\bf{Band 9}} \\ 
\hline
$<$3.7 &0.960  &  -0.052 $\pm$ 0.015\\ 
\hline
\hline
\end{tabular}
}
\label{tab6}
\end{center}
\end{table}

\begin{table}[ht!]
\begin{center}
  \caption{Linear fit parameters for the corrected in-band coherence against in-band calibrator separation angle separated by visibility baseline lengths. }
  {\footnotesize
\begin{tabular}{@{}lcc@{}}
\hline
Baseline  length ($\lambda$)   & Offset  &  Slope   \\ 
\hline
\hline
\multicolumn{3}{c}{\bf{Short}} \\   
\hline
\hline
$<$5$\times10^{6}$   & 0.954  & -0.022 $\pm$ 0.005\\ 
\hline
\hline
\multicolumn{3}{c}{\bf{Long}} \\ 
\hline
\hline
$>$5$\times10^{6}$ & 0.945 &  -0.054$\pm$ 0.012\\ 
\hline
\hline
\end{tabular}
}
\label{tab6_5}
\end{center}
\end{table}

Plotted differently, in Figure \ref{fig14} we split the data into two groups based on the baseline length in units of wavelength (to remove any frequency dependence). The top and bottom panels plot baselines shorter and longer than 5$\times$10$^{6}\lambda$, $\sim$5000\,m at our band 7 frequency. The divide therefore groups the short-baseline band 8 and mid-baseline band 7 data into the top panel and all the band 9 datasets, band 8 mid-baselines and band 7 long-baselines into the bottom one. Fitting each visibility baseline length group, following the exceptions as above, we find gradients of -0.022$\pm$0.005 for $<$5$\times$10$^6\lambda$ and -0.054$\pm$0.012 for $>$5$\times$10$^6\lambda$ (see also Table \ref{tab6_5}), clearly highlighting the baseline length dichotomy. The y-intercept values are again close to one, although the shorter baselines are still skewed by the two lower-coherence images, while the longer baselines are affected by the data spread as the fit combines band 7 and scattered band 9 values. The black solid line indicates the fit to a set of ideal observations with antenna position uncertainties added, we discuss this in Section \ref{hardlim}.

\begin{figure*}  
\begin{center}
  \includegraphics[width=18.5cm]{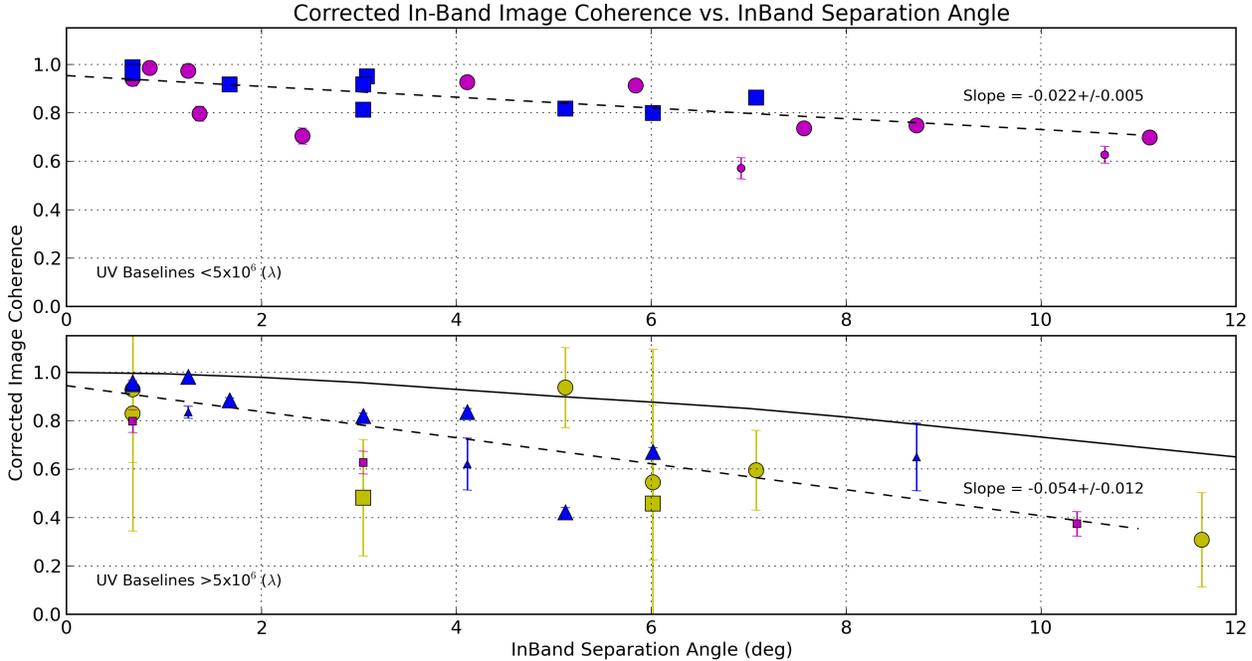} 
\caption{Corrected in-band image coherence as a function of in-band phase calibrator separation angle as Figure \ref{fig13} but with the datasets separated by visibility baseline length in units of wavelength. The top and bottom panels separate the maximal baselines that are shorter or longer than 5$\times$10$^{6}\lambda$ ($\sim$5000\,m at our band 7 frequency). The divide splits the data into two groups where different fitted slopes emerge for each. Slopes are fitted for the ensemble datasets in each visibility baseline length group using only those with low phase RMS (dashed lines). Symbols and Colors are the same as Figure \ref{fig2}. The solid line represents a simple model indicating the effect of only antenna position uncertainties. Errors are propagated from: an uncertainty in the peak flux density that is three times the image noise; and a variation of 20\% in the phase rms. In some cases these are smaller than the symbol.}
\label{fig14}  
\end{center}
\end{figure*}

\section{Discussion}
\label{disc}
One of the fundamental questions for ALMA is: which phase calibration technique should be used for high frequency observations? We can now begin to answer this based on the presented results.

\subsection{Degradation with separation angle}
\label{hardlim}

Results presented throughout Section \ref{calsep}, specifically in Section \ref{freqBL} and most clearly visualized in Figures \ref{fig13} and \ref{fig14}, strongly demonstrate that a small separation is required between the phase calibrator and the target source to minimize decoherence, assuming self-calibration is not possible. As a further example, Figure \ref{fig15} shows images of various point-source targets calibrated using standard in-band phase referencing where the calibrator to target separation angles are $>$5$^{\circ}$. Top, middle and bottom panels show bands 7, 8 and 9 images respectively. The visibility baseline lengths are $>$5$\times$10$^6\lambda$ for all expect the left-middle panel, which shows a slightly shorter baseline band 8 observation (max baseline $\sim$4.4$\times$10$^6\lambda$). The expected phase RMS values of these observations are lower than 30$^{\circ}$ and thus the expected image coherence should be $>$87\,\%, except the right-middle panel, which shows $\sim$40$^{\circ}$ and should have achieved a coherence of at least 78\,\%. The expected coherence levels are not met for any images. The reduced coherence and target structure defects seen in the images are solely due to the use of distant phase calibrators. We note that the lowest resolution (i.e. shortest baseline) observation (left-middle) is structurally the least degraded, as one also expects from the previous trends.

Considering the evidence provided, we know that shorter baselines and lower frequencies are less susceptible to degradation caused by distant calibrators, as expected. On the contrary, longer baselines and higher frequencies (independently and to a greater effect when combined) are most affected. Also, short-baseline ($\sim$3\,km) band 9 observations behave similar to long-baseline ($>$8.5\,km) band 7 ones, and thus should follow similar constraints for phase calibrator separation angles. If we consider the possible path length errors cause by antenna position uncertainties this behavior is somewhat expected, because these uncertainties scale with baseline length and with frequency when converted to phase (see also Section \ref{intro}). Using only the vertical-direction baseline dependent error (0.198\,mm/km, \citealt{Hunter2016}) we estimate the path length uncertainties $\ell$ ($\mu$m) via $\Delta \rho \cdot (s_t - s_c)$. Here $\Delta \rho$ is the baseline position uncertainty and $(s_t - s_c)$ is the calibrator to target separation angle (in radians). For baseline lengths of 15, 10 and 5\,km, $\vert\Delta\rho\vert$ = 2.97, 1.98 and 0.99\,mm respectively, and thus for a calibrator to target angle of 5$^{\circ}$, $\ell\sim$ 259, 173 and 86\,$\mu$m. During a relatively short observation the projected baselines and the on-sky target-to-phase calibrator separation angle will remain roughly constant. Therefore, via Equation \ref{eqn0}, at band 7, 8 and 9 for the long-, mid- and short-baselines almost constant phase offsets of 90, 84 and 68$^{\circ}$, respectively, would be imparted to the target during phase referencing. These values are reasonably similar, illustrating our expected result tying low-frequency longer baselines to high-frequency shorter baselines. The expected coherence would range from 30-50\,\% for phase RMS values of this magnitude (phase RMS = phase offset for data with a zero degree mean phase) before we begin to consider the effects of atmospheric variations that remain after phase referencing. This simplistic treatment using a single maximal baseline length underestimates the expected coherence when compared with the data, because many shorter baselines in a real array would not suffer such a large corruption.

To provide a more realistic case we use the self-calibrated data of the DGC source from the long-baseline J0449-170928-B37-2deg dataset and corrupt it with incorrect antenna positions generated using {\sc gencal} in {\sc casa} for a range of calibrator separation angles, from 1 to 12$^{\circ}$ (in 1$^{\circ}$ steps). We provide the antenna position uncertainties as an input, calculated with the same method as above, for all baselines made with the reference antenna but in all directions, east, north and vertically. After refitting the data presented in \citet{Hunter2016} we use the uncertainties as a function of baseline length, $b$, of: east (0.140\,mm +  $b$\,km\,0.071\,mm/km); north (0.110\,mm +  $b$\,km\,0.054\,mm/km); and vertical (0.220\,mm + $b$\,km\,0.198\,mm/km). The baseline-length-independent position errors are now accounted for. We randomly select a final position uncertainty for each direction from a Gaussian distribution centered on the absolute uncertainty as above, per baseline, with standard deviations in the east, north and vertical directions of 0.086, 0.076 and 0.129\,mm and 0.121, 0.118 and 0.296\,mm for baselines shorter and longer than 2.5\,km, respectively. These values were found using the data from \citet{Hunter2016} after removing the baseline-dependent fit to find the spread. The signs of the position uncertainties are randomly assigned for each antenna. A Gaussian phase variation of 20$^{\circ}$ is added thereafter to each baseline, representative of that expected to remain in well calibrated observations.

\begin{figure*}  
\begin{center}
  \includegraphics[width=18.5cm]{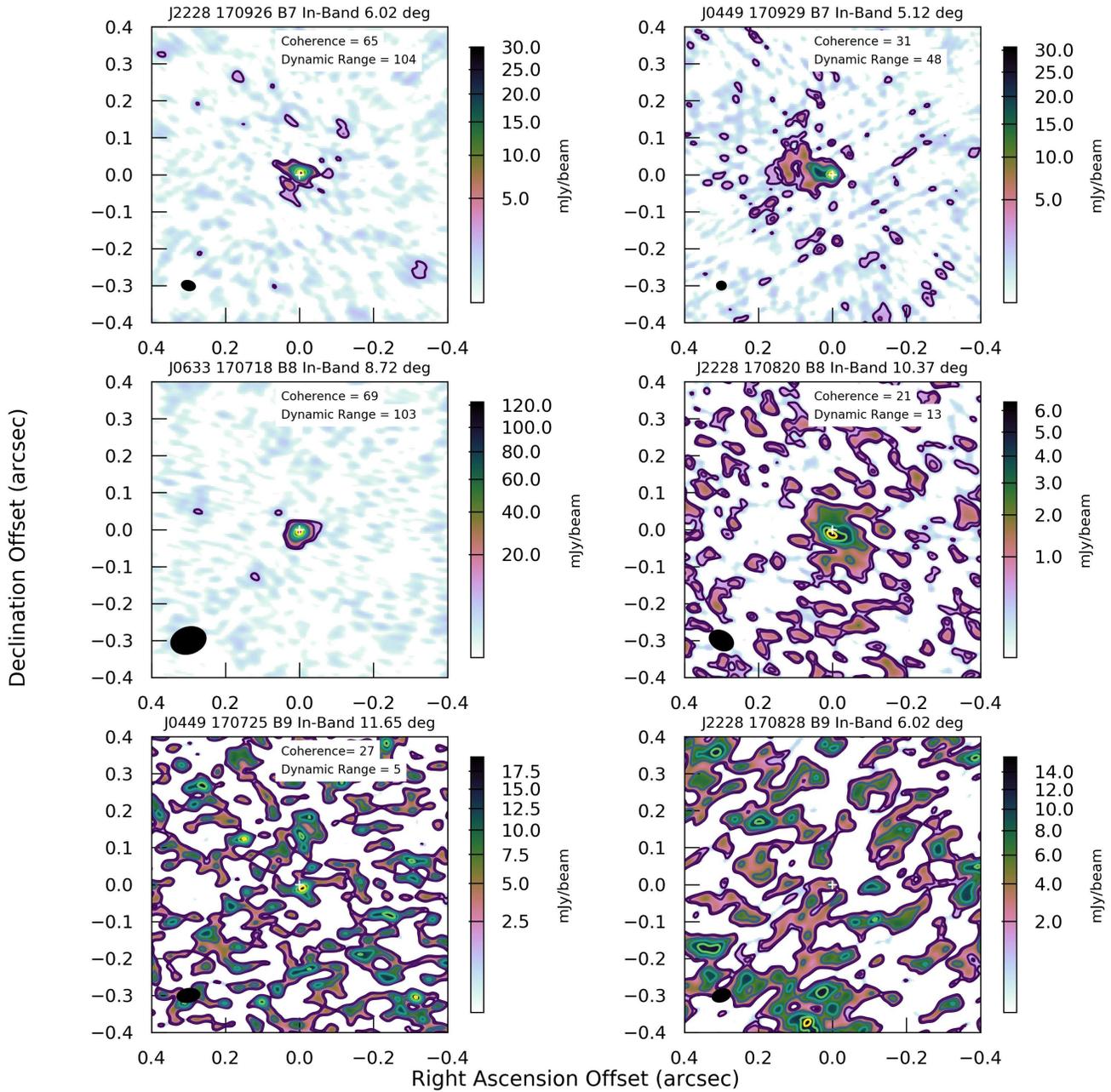}
\caption{Images of various target sources calibrated using standard in-band phase referencing but where phase calibrators have an angular separation $>$5$^{\circ}$ Top-left to bottom right are images of the targets J2228-0753, J0449-4350, J0633-2223, J2228-0753, J0449-4350 and J2228-0753. The top, middle and bottom panels show bands 7, 8 and 9 respectively. The images are all scaled to the respective peak flux values, and the contours are plotted at the 5, 10, 30, 50, 70 and 90\,\% levels. The panel titles identify the observation and the calibrator separation angle. The beams are shown as a black ellipses to the bottom left, while the coherence (\%) and dynamic range are shown at the top right. The expected phase RMS for all observation was $<$30$^{\circ}$ and the images should have achieved image coherence valves $>$87\,\%, except the right-middle panel band 8 image which was $\sim$40$^{\circ}$ corresponding to an image coherence of 78\,\%. All images have a coherence lower than expected and some have considerable structural defects. The white cross marks the center of the image. We note that J2228-0753 is not imaged in band 9 (bottom-right) and no parameters are shown.}
\label{fig15} 
\end{center}
\end{figure*}

The solid black line in the lower panel of Figure \ref{fig14} shows the fit to the corrected image coherence of the modeled observations corrupted by antenna position uncertainties. We apply the same correction as used for the observed data, adding the expected coherence loss due to atmospheric phase variations to the measured image coherence. The correction is 6\,\% for all model images as we corrupted the data with only a 20$^{\circ}$ phase RMS. As expected, for zero separation angle there are no antenna position uncertainties and the correction applied for the atmospheric phase RMS sets the image coherence to one. The degradation of the antenna-position corrupted models follows a curved fit (corrected coherence = -0.00135\,$\theta^2$ -0.01338\,$\theta$ $+$ 1.00638, where $\theta$ is the separation angle in degrees). The fit has a noticeably shallower decline of corrected image coherence when compared with the long visibility baseline data ($>$5$\times$10$^6\lambda$). At separation angles of 2, 4 and 6$^{\circ}$, the antenna position errors amount to image coherence losses of $\sim$3, 7 and 12\,\%, respectively. The corrected image coherence values, with only antenna position corruptions are 13, 20 and 36\,\% higher than the fit to the observations. Antenna position uncertainties therefore cannot fully account for the level of image decoherence. Considering a case where antenna position uncertainties were negligible, assuming we can naively add the decoherence they cause to the observational data, then a re-fit to the observations with antenna position uncertainties removed yields a gradient of -0.026$\pm$0.012. If we can correct for antenna position uncertainties, even partially, degradation as a function of separation angle can be significantly reduced, particularly for larger separation angles. 

The cause of the remaining decoherence, over that attributable to antenna position uncertainties, is sub-optimal phase referencing. A previous investigation of ALMA band 3 long-baseline observations conducted by \citet{Asaki2016} reported that the phase RMS for targets at 5.0$^{\circ}$ and 5.5$^{\circ}$ were not significantly improved after phase referencing, even when using fast switching, which should have corrected any rapid atmospheric fluctuations. Similarly, poor phase correction was illustrated at millimeter wavelengths in paired antenna tests by \citet{Asaki1998} using the NMA with baselines $<$1\,km. Their study used a satellite as a calibrator source such that the separation angle between it and the celestial target varied from 0 to 50$^{\circ}$ during the observations. With the paired antenna system the target and satellite were continuously monitored. The difference of the phase-time series for the target and satellite indicated that effective phase correction occurred only when the satellite was within $\sim$10$^{\circ}$ of the target, and that only if it was within $\sim$2$^{\circ}$ was the path length after correction below 100\,$\mu$m. This path length was a requirement at the time of the study in which future observatories could successfully undertake sub-mm observations. Beyond a 10-15$^{\circ}$ angular separation there was no correction to the target phases. Interestingly, \citet{Asaki1998} were able to better phase calibrate their observations via the introduction of a time-lag into the phase-time series as a function of increasing separation angle which compensated for the differing target and calibrator lines-of-sight as the atmosphere advected over the array. It is also worth noting a study of calibrator separation angles using VLBI data at 8.4 and 15.0\,GHz frequencies by \citet{MartiVidal2010}. In short, these authors also find a clear decrease of image coherence (they call this peak-ratio) with separation angle. Their results also tie with their previous modeling of a snapshot turbulent atmosphere, which included both ionospheric and tropospheric effects, the former important at such low frequencies \citep{MartiVidal2010b}. Unlike the ALMA observations we present, those of \citet{MartiVidal2010} do not further deteriorate past separation angles of $\sim$15$^{\circ}$ where is it believed that the atmospheric turbulence saturates at these lower observing frequencies. For high-frequencies, our findings combined with those of \citet{Asaki1998, Asaki2016} imply that close calibrators are required to correctly track and correct the fluctuations near to the line-of-sight of a target. 

In Figure \ref{fig16} we indicate the residual phase RMS of the target sources pre- and post- phase calibration from all band 7 and 8 observations that should have achieved an expected phase RMS $<$30$^{\circ}$. The symbol sizes are indicative of the target-to-calibrator separation angles, with dark and light symbols representing in-band and B2B calibration. The B2B data all use close calibrators. We remove any nonzero phase offsets from the phase RMS to exclude the effect of antenna position uncertainties that manifest as almost constant phase offsets in time, and thus the residual phase RMS values plotted are assumed to be purely due to uncorrected atmospheric fluctuations.  Even though our observations were taken in a variety of stability conditions, indicated by the large spread of pre-calibration phase RMS values, the targets using nearby calibrators are generally corrected to, or below, the expected phase RMS level (30$^{\circ}$). The post-calibration residual phase RMS values for in-band data using distant calibrators generally show values worse than the expect phase RMS level. For some in-band data with distant calibrators that do meet the expected level, they already had low pre-calibration phase RMS values and are actually corrected very little by phase referencing (e.g. larger blue-squares near a pre-calibration phase RMS of $\sim$30-40$^{\circ}$). In-band data with distance calibrators therefore indicate sub-optimal calibration.

Crucially, when using distant phase calibrators, in addition to antenna position uncertainties, uncorrected atmospheric phase variations still remain in data that should have been corrected to a much lower phase RMS had a close calibrator been used. 

\begin{figure}  
\begin{center}
  \includegraphics[width=8.5cm]{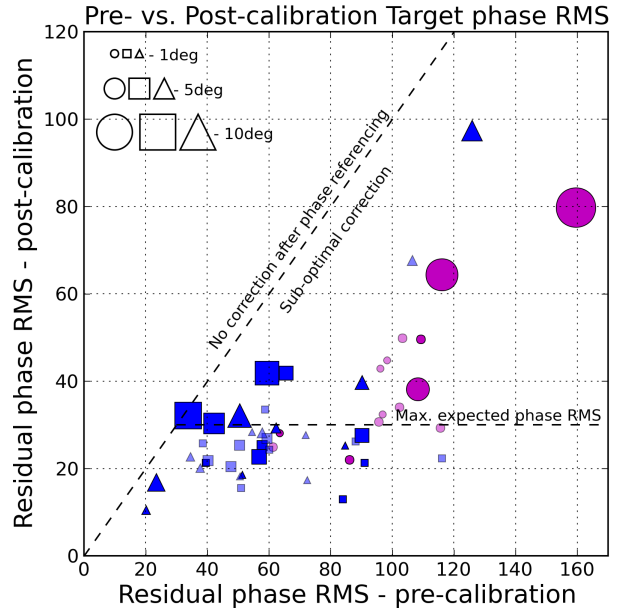} 
\caption{Pre- and post-calibration residual phase RMS of the target sources from all band 7 and 8 observations that were expected to achieve a low phase RMS ($<$30$^{\circ}$). Colors (bands) and symbols (maximal baseline length) follow Figure \ref{fig2}, while symbol size is indicative of phase calibrator separation angle and dark and light colors are for in-band and B2B observations respectively. The diagonal dashed line mark regions of no correction (left) and sub-optimal correction (right) while the horizontal one marks the line of expected maximal phase RMS (30$^{\circ}$). Most B2B datasets with close calibrators have post-calibration residual phase RMS values that meet the expected phase RMS, whereas many in-band datasets fall in the region of sub-optimal phase correction due to using distant calibrators. Some in-band data with distant calibrators do meet the expect phase RMS level, however, these already had very low pre-calibration phase RMS values (e.g. large blue-squares near $\sim$30-40$^{\circ}$ pre-calibration phase RMS).}
\label{fig16}
\end{center}
\end{figure}

\subsection{Recommended maximal separation angles}
\label{reclim}
If a maximal calibrator separation angle is not employed images could be degraded to a level where they would negatively impact any scientific interpretation. There could also be a significant impact for detection experiments where the reduced coherence for weak targets would reduce peak flux densities below suitable confidence thresholds. The images presented in Figure \ref{fig15} indicate coherence values below $\sim$70\,\% are associated with an increase of structural defects when the calibrator separation angles exceed $\sim$5$^{\circ}$. With this as a reference point and following the empirical fits to Figure \ref{fig14} provided in Table \ref{tab6} we parameterize the recommend maximal phase calibrator to target separation angle required to achieved a minimal 70\,\% image coherence. As noted in Section \ref{freqBL}, the final image coherence has two components: the expected coherence derived from the expected phase RMS; and the additional coherence loss due to separation angle. To reiterate, the former parameterizes the highest achievable coherence tied to the lowest expected phase RMS after phase referencing with an ideal zero degree separation angle. The latter encompasses antenna position uncertainties and any sub-optimal phase correction of the temporal atmospheric fluctuations due to the finite calibrator-to-target separation angle. Following the divide used throughout this work, we assume that atmospheric fluctuations can be corrected to an expected phase RMS of 30$^{\circ}$, and thus the maximal image coherence should be 87\,\%, a loss of only 13\,\%. In evaluating the separation angle required to achieve the final image coherence of 70\,\%, the expected coherence loss is subtracted from the y-intercept of the fits presented in Table \ref{tab6}. The maximal separation angles using the shallowest slope limit and maximal baseline length of each group are presented in Table \ref{tab7}. We note that these limits need not be as strict for better observing conditions where the expected phase RMS would be $<$30$^{\circ}$ as the coherence loss would be $<$13\,\%, or for sources that can be self-calibrated. However, when considering self-calibration image structural defects would still need to be minimized for targets with complex extended structure. We also fit a planar function to these maximal recommended separation angles in order to fill the remaining baseline and frequency ranges that were not covered in our test observations. We caution that the planar fit uses only four data points. For band 9 long-baselines the value is un-physical (negative), which may point to a real difficulty in achieving a high enough image coherence or possibly that the linear fit used to parameterize the image coherence degradation with separation angle is too simplistic. 

\citet{Asaki2020} presented a table listing the mean separation angles of suitable calibrators for high-frequency observations. Those authors indicated that for bands 9 and 10 in-band calibrators would be $>$7.8$^{\circ}$ away while B2B calibrators are almost a factor of two closer at $\sim$4.1$^{\circ}$. Certainly B2B appears to be the only foreseeable way to calibrate high-frequency observations. The B2B average separation angle of available calibrators is larger than our requirement for band 8 long-baselines in Table \ref{tab7} and suggests higher frequency observations may not achieve a minimal 70\,\% coherence level. In the present long-baseline observing regime, where ALMA uses 72\,s cycle times with $\sim$18\,s spent on the calibrator, successful high-frequency observations may only be conducted if the science target is luckily within 1-2$^{\circ}$ of a strong quasar. Alternatively if the atmosphere was stable enough, longer cycle-times with longer on-calibrator times could be used to attain suitably high S/N solutions for weaker, but close, calibrators. Following from the discussion in Section \ref{hardlim}, it is apparent that other remedial action could be made. The effect of antenna position uncertainties could be minimized by performing a short observation (at a lower frequency) targeting a number of quasars within a few degrees of the target in order to calculate the vertical (zenith) path length errors. A better correction of these at the time of observing would minimize the effect of the vertical uncertainties in the antenna positions and could increase the calibrator separation angle tolerance to match the B2B calibrator availability. It is worth noting that long time-interval self-calibration could also provide some correction for uncertain antenna positions which are manifest as almost constant phase offsets. However, a detailed study with high frequency long-baselines observations is required to understand the divide between sub-optimal phase referencing and antenna position errors as a function of calibrator separation angle, and whether long term self-calibration is feasible. 

\begin{table}[ht!]
\begin{center}
  \caption{Maximum recommended calibrator separation angle to achieve an image coherence of 70\,\%.}
  {\footnotesize
\begin{tabular}{@{}lccc@{}}
\hline
 Band (Frequency)   &     $b$$<$3.7\,km    &  $b$=3.7-8.5\,km        &  $b$=8.5-16.0\,km   \\ 
\hline
\hline
7 (279\,GHz)    &  10.5$^a$   &   8.4   &   3.8   \\  
8 (394\,GHz)    &      8.2   &      6.0$^a$   &   2.0$^a$         \\ 
9 (681\,GHz)    &         3.7       &       1.0$^a$          &    - $^b$\\
\hline
\hline
\end{tabular}
}
\begin{tablenotes}
Notes: The separation is estimated using the lower-limit gradient in Table \ref{tab6} for various baseline lengths, $b$, assuming an expected phase RMS of 30$^{\circ}$, attributable to a 13\,\% coherence loss.\\
$^a$Estimated using a planar fit. \\ 
$^b$ $-$ Unphysical value.
\end{tablenotes}
\label{tab7}
\end{center}
\end{table}

\subsection{Choice of technique}
\label{discchoose}
Given that we have now established suitable maximal limits for the calibrator to target separation angle, we must ask: when does B2B deliver higher image coherence than in-band when calibrators are found for both?

To make this assessment we compare the difference between in-band and B2B image coherence values for the paired observation as a function of the difference in the phase calibrator separation angles. We subtract the B2B image coherence from the in-band values. In cases where the same phase calibrators were used for B2B and in-band, the difference in separation angle will be zero degrees, while if the in-band images have a higher coherence, the coherence difference will be positive. The in-band data differenced with the B2B data are plotted as filled symbols in Figure \ref{fig17}. Many of the test observations were taken consecutively on the same day, of which the in-band blocks often observe the same target source in the same band but with a different phase calibrator. We have 28 additional unique pairings of in-band only observations from which we plot the image coherence of the observation after differencing with that using the closer calibrator for the various pairings. These in-band data are plotted as open symbols in Figure \ref{fig17}. 

\begin{figure*}  
\begin{center}
  \includegraphics[width=18.5cm]{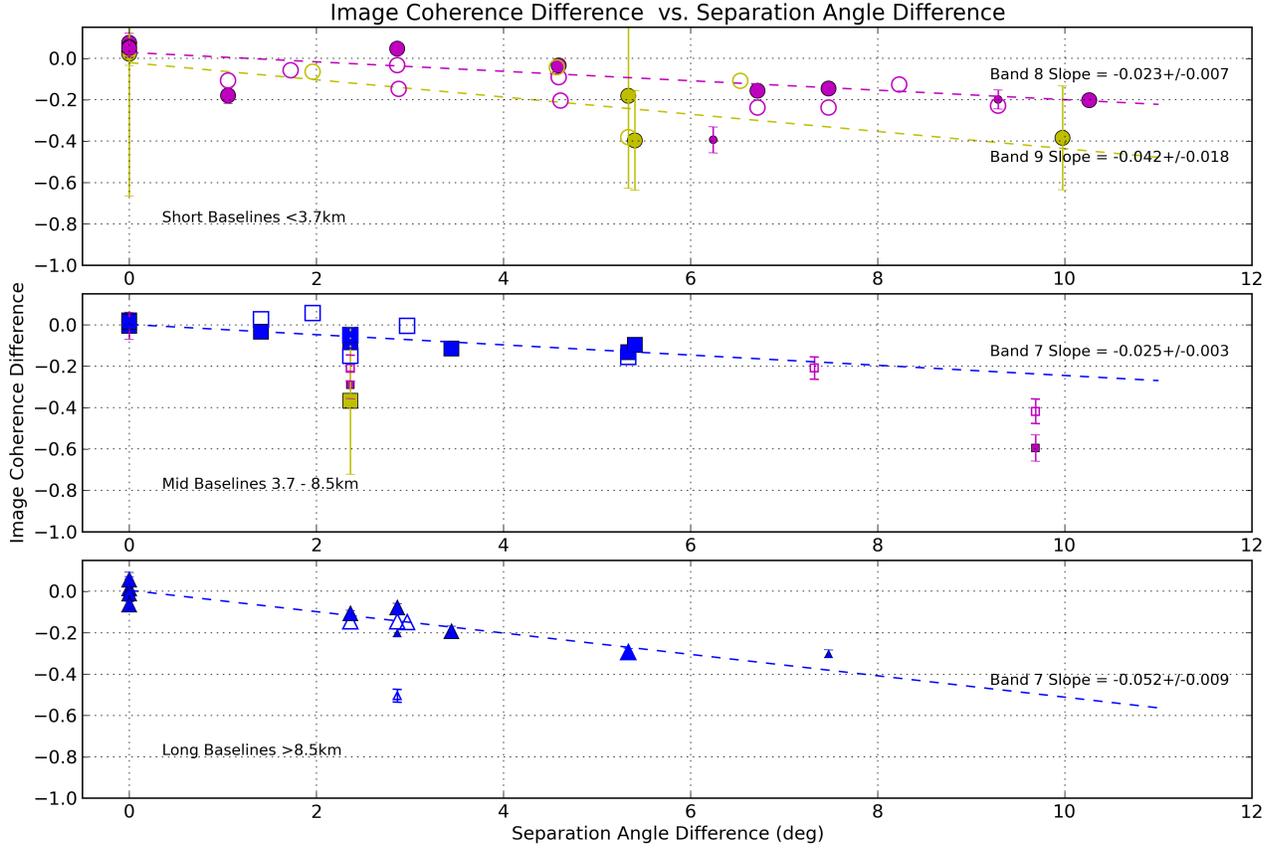} 
\caption{Image coherence difference resulting from the subtraction of the B2B image coherence from the in-band paired observations (filled) and from in-band observations paired with other in-band observations on the same day but using a different calibrator (open) as a function of phase calibrator separation angle difference, for short- ($<$3.7\,km, top), mid- (3.7$-$8.5\,km, middle) and long- ($>$8.5\,km, bottom) baseline length array configurations. All datasets are plotted including those where the in-band and B2B blocks use the same phase calibrators, i.e. at zero degrees separation angle difference. There is a notable image coherence loss as a function of calibrator separation angle difference. The larger of the two symbol sizes highlight those datasets where the expected phase RMS, as measured over a $\sim$30\,s interval, are $<$30$^{\circ}$. The dashed lines are fits using only low expected phase RMS data for the different frequencies and baseline lengths, and using only in-band data differenced with B2B data. Table \ref{tab8} presents the fit results. Colors (bands) and symbols (maximal baseline length) are the same as in Figure \ref{fig2}. Band 9 observations with much poorer B2B images are off the scale of the top panel and are excluded in the fitting. Clearly using a closer calibrator will result in a better image coherence.}
\label{fig17} 
\end{center}
\end{figure*}

\begin{table}[ht!]
\begin{center}
  \caption{Linear fit parameters from the image coherence difference against calibrator separation angle difference. }
  {\footnotesize
\begin{tabular}{@{}lcc@{}}
\hline
Baseline  length (km)   & Offset  &  Slope   \\ 
\hline
\hline
\multicolumn{3}{c}{\bf{Band 7}} \\   
\hline
\hline
3.7-8.5   & 0.003   & -0.025 $\pm$ 0.003\\%
$>$8.5  & 0.006  &  -0.052 $\pm$ 0.009\\%
\hline
\hline
\multicolumn{3}{c}{\bf{Band 8 }} \\ 
\hline
\hline
$<$3.7 & 0.030  &  -0.023$\pm$ 0.007\\  
\hline
\hline
\multicolumn{3}{c}{\bf{Band 9}} \\ 
\hline
$<$3.7 & -0.019 & -0.042 $\pm$ 0.018\\ %
\hline
\hline
\end{tabular}
}
\label{tab8}
\end{center}
\end{table}

Figure \ref{fig17} can be viewed as directly indicating a further degradation of in-band observations when specifically compared to B2B observations independent of the absolute calibrator separation angle. In principle if in-band and B2B observations used the same calibrator, be it at an absolute separation angle of 2$^{\circ}$ or 5$^{\circ}$, both would be plotted at a separation angle difference of 0$^{\circ}$ and thus the coherence difference would be around zero. Evidently there is a notable image coherence degradation as a function of phase calibrator separation difference. These trends can be used to decide if the in-band or B2B technique should be preferred in returning the highest coherence images. Simply, using a linear fit (see Table \ref{tab8}) we can evaluate $m\Delta\theta + k$, where $k$ is the y-axis intercept, $m$ is the gradient and $\Delta\theta$ is the angular difference (in degrees) between in-band and B2B calibrators. Values $>$0 mean that in-band phase referencing would be selected. Indeed, at zero degrees difference (i.e. the same calibrator) for all baselines and bands 7 and 8 the intercept values, $k$, are $>$0 and in-band would always be used. We note that band 9 data has a negative intercept, however the fit only uses 4 datasets each with considerable uncertainties. As soon as $m\Delta\theta + k < 0$, the B2B technique would be preferred to provide a calibration yielding the highest coherence image. For the band 7 long-baseline data investigated this would already occur when a B2B calibrator was only $\sim$0.12$^{\circ}$ closer to the target compared to the available in-band one. 

Interestingly, the y-axis intercept values for the band 7 mid- and long-baselines groups are almost zero, suggesting that the B2B images are not noticeably degraded due to the added DGC step in comparison with the band 8 observations. This however is a side effect of the data at hand. The band 7 and band 8 datasets with low expected phase RMS have average residual DGC source phase RMS values of $\sim$13$^{\circ}$ and $\sim$28$^{\circ}$, respectively. With reference to Section \ref{dgcdirect} the DGC detrimental effect on the band 7 data is $<$1\,\%, but is $\sim$6\,\% for the band 8 data. The net effect of the DGC for our presented data is to change the y-axis intercept of the fits presented in Table \ref{tab8}, and thus should be considered with care. This is discussed further in Section \ref{metric}.
 
\subsection{Defining an operational metric}
\label{metric}
From an operations perspective we have to take a pragmatic approach, and not just consider which technique provides the highest coherence image judged from the linear trends presented in Section \ref{discchoose}. Time overheads for the B2B mode must be accounted for, and likewise the detrimental effect of the DGC considering the permissible observing conditions. The operation metric can be broken into three parts: inaccuracies in the DGC solution, time overheads, and which technique provides the best image coherence for the phase calibrators available.

Firstly, following the previous scenario we consider the case where the expected phase RMS remaining after phase referencing is 30$^{\circ}$ for the DGC source. Therefore, we could expect that the detrimental effect of DGC would reduce the coherence of a B2B image by up to 7\,\%, see also Section \ref{dgcdirect}. Indeed, if the expected phase RMS could be reduced further, by observing in better conditions, the detriment to the B2B target image will become almost negligible.

Secondly, observations at ALMA are broken up into execution blocks, EBs. Each EB lasts a pre-defined amount of time, limited by the maximal target source integration time of 50\,min, and is a self-contained dataset with all required calibrators. In order to achieve the scientific goal of any project the EB is repeated until the sensitivity requirement is met after calibration and imaging. For B2B projects a DGC source would be added to the EB. If observations follow a similar regime to the presented tests the DGC source would be observed for $\sim$5\,min at the start and end of each EB. In an attempt to incorporate the time overhead into the operational metric we consider the case of a fixed length EB where the maximal in-band on-source time would be 50\,min. For B2B mode we thereby subtract 10\,min on-target time, reducing it to 40\,min. Consequently, as the image map noise is proportional to the square root of time on source, $\Delta\sigma_{\rm image} \propto 1.0/\sqrt{\Delta t_{\rm obs}}$, the B2B image noise level would increase to $1.0/\sqrt{0.8}$ = 1.12 of the in-band value for an EB of the same total duration. An increase of 12\,\%. However, in Section \ref{conds} we identified that the in-band image map noise when using distant calibrators is typically larger than the B2B images using closer ones. There did not appear to be a definitive trend with separation angle, thus excluding images where the in-band noise is $>$30\,\% worse than B2B images, the average noise increase for in-band data is 8\,\%. Overall balancing the time overhead of B2B as a noise increase with the in-band noise increase due to using distant calibrators, there is only a 4\,\% deficit for B2B. We note that in reality the B2B observations will be extended in duration to ensure that the noise requirements are met, and that the above calculation is an illustrative way to account for the cost, in terms of image parameters, of the extra time. 

The third and final term was already found in Section \ref{discchoose}. The gradient for the fits in Figure \ref{fig17} indicates the extra image degradation per degree of calibrator separation angle difference. We note that the gradient is the only negative term of the operational metric. As a reminder, the y-axis intercept is not used from those fits, as it is biased by the low residual phase RMS of the DGC source of those data. We address this by instead using the first term of the metric described above. 

The three terms of the operational metric, in order, can now be combined as shown in Equation \ref{eqn3}:

\begin{equation}
\label{eqn3}
0.11 + m\Delta\theta  = \left\{   
\begin{array}{rl}
{\rm B2B} & {\rm if} < 0\\
{\rm in-band} & {\rm if} > 0
\end{array}\right.
\end{equation}

\noindent where the constant, 0.11, combines the effects from terms one and two (as a fractional value), $m$ is the gradient of the fits from Table \ref{tab8} dependent on baseline length and frequency and $\Delta\theta$ is the angular difference (in degrees) between in-band and B2B calibrators. Following the same rational as Section \ref{discchoose}, B2B would be applied in operations based on the evaluation of Equation \ref{eqn3} being $<$0, otherwise a standard in-band observation should be used under the premise that the maximal recommended calibrator separation angle is not exceeded. To provide two clear cases relevant for the upcoming ALMA band 7 long-baseline observations: if an in-band calibrator was found only beyond the maximal separation angle limit of 3.8$^{\circ}$ then B2B would be the default technique used; whereas if an in-band calibrator was found at the maximal limit from the target, B2B would only be employed if the calibrator is 2.1$^{\circ}$ closer (i.e. $<$1.7$^{\circ}$ from the target). 

\section{Summary}
\label{sum}
An investigation was made using 44 analyzed datasets taken as part of the ALMA high-frequency long-baseline campaign 2017 (HF-LBC-2017) in order to make a direct comparison between standard in-band phase referencing and band-to-band (B2B) phase referencing techniques. The B2B technique is a method to calibrate high frequency observations using phase solutions from a calibrator observed at a lower frequency, as the quasar calibrator is likely to be brighter and also closer to the target. Calibration involves correction of the instrumental phase offset between the frequencies and the conversion of the calibrator temporal phases to the frequency of the target. A differential-gain-calibration (DGC) sequence, consisting of alternating low and high frequency scans of strong quasar, is used to calibrate the instrumental offset. The test observations included both B2B and in-band phase referencing blocks which observed the same target sources that were calibrated with either the same, or different phase calibrators. A range of maximal baseline lengths were investigated from 2\,km out to $\sim$15\,km. Successful high frequency observations were made in bands 7, 8 and 9, which were paired with band 3, 4 and 4 and 6 respectively for the B2B mode (B7-3, B8-4, B9-4, B9-6). The work presented also examined the detrimental effects of increasing target to phase calibrator separation angles on calibration and imaging. 

In 16 observations the same nearby phase calibrator was chosen for the in-band and B2B blocks (separation angles $<$1.67$^{\circ}$). Comparing target image parameters such as peak flux density and image coherence we find the B2B calibration technique can produce images with similar properties, within 5-10\,\%, of the standard in-band phase calibration, the latter providing typically better images. The DGC step required for the B2B technique is responsible for the few percent difference between the B2B and in-band images. Provided that the phase residuals are minimized ($<$30$^{\circ}$) for the DGC source then the detriment to B2B images is $<$7\,\% in terms of coherence. This is reduced to $<$4\,\% for phase residuals $<$20$^{\circ}$.

For the remaining 28 datasets the phase calibrators for the B2B block were selected to be nearby ($<$1.67$^{\circ}$) whereas the in-band calibrators were chosen to be at a larger separation angles, between 2.42 and 11.65$^{\circ}$. These observations were designed to test the specific use case of B2B in which low frequency calibrators are much more likely to be found closer to a given target compared to finding a suitably strong calibrator at the in-band high frequency. Comparing the B2B and in-band images we find that the peak flux density, noise level and image coherence parameters are superior for the B2B images. Most in-band images have coherence values that are $>$15\,\% worse than the B2B ones. 

Examination of the in-band image coherence shows a linear decrease with increasing calibrator separation angles. The image coherence values are lower than the expected image coherence values which are calculated using the expected phase RMS measured over the cycle time as a proxy. Corruption of an ideal self-calibrated long-baseline observation with antenna position uncertainty effects can only partially account for the decoherence. The remaining coherence loss is attributed to sub-optimal phase referencing due to the different lines-of-sight through the variable atmosphere for the target and calibrator, respectively. Investigation of the target source RMS pre- and post-calibration residual phase RMS indicates that distant calibrators do not correct the phase fluctuations down to the expected level.

The trends of decreasing coherence with separation angle are baseline and frequency dependent. For our tested parameter space, long-baselines and high frequency observations (independently) show the largest image coherence degradation. Long-baseline band 7 observations ($>$8.5\,km) show a similar trend as short-baseline ($<$3.7\,km) band 9 observations. Making an ensemble comparison of all observations, divided in visibility baseline length at 5$\times$10$^6$\,$\lambda$, we find that shorter baselines have a fractional image coherence degradation of $-0.022\pm$0.005 per degree, while longer baselines show a steeper degradation of $-0.054\pm$0.012 per degree. 

Images made using long-baseline and calibrators exceeding $>$5\,deg typically have coherence values below 70\,\%, even when the expected coherence was $>$87\,\%. Notable image deformation and structural changes also occur and could effect a scientific interpretation. For band 7 long-baseline observations, assuming conditions that could achieve an expected $\sim$30$^{\circ}$ phase RMS corresponding to a $\sim$87\,\% coherence, a phase calibrator must be within $\sim$4$^{\circ}$ to actually achieve a final image coherence $>$70\,\%. Propagation to higher frequencies suggests that calibrators may need to be closer than 2$^{\circ}$ to a target, something that may only be possible to achieve using B2B observations.

Finally, we present a metric that could be used to judge whether the B2B mode should be used in operations depending on calibrator availability. The B2B mode should always be used if there is no suitably close in-band calibrator that would provide a final target image coherence $>$70\,\%. However, if calibrators are found for both techniques, the metric accounts for the effect of DGC on the B2B image, the time overhead required for the B2B mode, and the extra coherence degradation of in-band images compared to B2B images as a function of the difference in calibrator separation angle. For long-baselines $>$8.5\,km band 7 observations B2B would only be preferred if it could offer a calibrator $\sim$2$^{\circ}$ closer than the best in-band one.

\begin{acknowledgements}
ALMA is a partnership of ESO (representing its member states), NSF (USA) and NINS (Japan), together with NRC (Canada) and NSC and ASIAA (Taiwan), and KASI (Republic of Korea), in cooperation with the Republic of Chile. The Joint ALMA Observatory is operated by ESO, AUI/NRAO, and NAOJ. The authors thank the referee for their positive report. We would also like to thank all the Joint ALMA Observatory staff in Chile for performing the challenging HF-LBC-2017 successfully, and Todd Hunter for discussions regarding antenna position uncertainties and SMA dual-frequency observations. L. T. Maud was adopted as an JAO ALMA expert visitor during his stays at JAO during the campaign. This work was supported by JSPS KAKENHI Grant Number JP16K05306. SM is supported by the Ministry of Science and Technology (MOST) in Taiwan, MOST 107-2119-M-001-020 and MOST 108-2112-M-001-048.
\end{acknowledgements}

\bibliographystyle{aasjournal}

\end{document}